\documentclass[manuscript,screen]{acmart}
\AtBeginDocument{%
	\providecommand\BibTeX{{%
			\normalfont B\kern-0.5em{\scshape i\kern-0.25em b}\kern-0.8em\TeX}}}
\usepackage[noend]{algpseudocode}
\usepackage{graphicx}
\usepackage{caption}
\usepackage{subcaption}
\usepackage{amsmath}
\usepackage{todonotes}
\usepackage{xspace}
\usepackage{microtype}
\usepackage{color}
\usepackage{ifthen}
\usepackage{pgf,tikz,pgfplots}
\usepackage{tikz-3dplot}
\usepackage{tkz-graph}
\usepackage{scalerel}
\usepackage{algorithmicx}
\usepackage{algorithm}
\usepackage{algpseudocode}
\usepackage{comment}
\usepackage{float}
\usepackage{placeins}
\usepackage{longtable}

\usepackage[tableposition=below]{caption}
\newtheorem{theorem}{Theorem}

\newtheorem{definition}[theorem]{Definition}

\newcommand{\Pol}{{\mathcal P}}
\newcommand{\subscripttriangle}{{\scaleto{\triangle}{5pt}}}
\newcommand{\triangleone}{\dot\triangle}
\newcommand{\triangletwo}{\ddot\triangle}
\newcommand{\subscripttriangleone}{\dot\subscripttriangle}
\newcommand{\subscripttriangletwo}{\ddot\subscripttriangle}

\newcommand{\ignore}[1]{}

\newboolean{formula}
\setboolean{formula}{false}

\date{}
\usetikzlibrary{calc,quotes,angles}

\begin{document}

	\title{Computing Area-Optimal Simple Polygonizations}%
	\author{S\'andor P.~Fekete}%
	\email{s.fekete@tu-bs.de}%
	\orcid{0000-0002-9062-4241}
	\affiliation{
		\institution{{Department} of Computer Science, TU Braunschweig}
		\city{Braunschweig}
		\country{Germany}
	}
	\author{Andreas Haas}
	\email{a.haas@tu-bs.de}
	\affiliation{
		\institution{{Department} of Computer Science, TU Braunschweig}
		\city{Braunschweig}
		\country{Germany}
	}
	\author{Phillip Keldenich}
	\email{p.keldenich@tu-bs.de}
	\orcid{0000-0002-6677-5090}
	\affiliation{
		\institution{{Department} of Computer Science, TU Braunschweig}
		\city{Braunschweig}
		\country{Germany}
	}
	\author{Michael Perk}
	\email{m.perk@tu-bs.de}
	\orcid{0000-0002-0141-8594}
	\affiliation{
		\institution{{Department} of Computer Science, TU Braunschweig}
		\city{Braunschweig}
		\country{Germany}
	}
	\author{Arne Schmidt}
	\email{arne.schmidt@tu-bs.de}
	\orcid{0000-0001-8950-3963}
	\affiliation{
		\institution{{Department} of Computer Science, TU Braunschweig}
		\city{Braunschweig}
		\country{Germany}
	}
	\begin{abstract}
	We consider methods for finding a simple polygon of minimum ({\sc Min-Area}) or maximum
	({\sc Max-Area}) possible area for a given set of points in the plane. Both problems
	are known to be NP-hard; at the center of the recent CG Challenge,
	practical methods have received considerable attention. However, previous methods
	focused on heuristic methods, with no proof of optimality.  We develop exact
	methods, based on a combination of geometry and integer programming.
	As a result, we are able to solve instances of up to $n=25$ points to provable
	optimality. While this extends the range of solvable instances by a considerable amount,
	it also illustrates the practical difficulty of both problem variants.
	\end{abstract}

\setcopyright{acmcopyright}

\acmJournal{TALG}
\acmVolume{99}
\acmNumber{99}
\acmArticle{99}
\acmMonth{13}



\maketitle
	\section{Introduction}
\label{sec:introduction}

\begin{figure}[h]
  \centering
    \includegraphics[trim={25mm 16mm 21mm 16mm},clip,width=.35\textwidth]{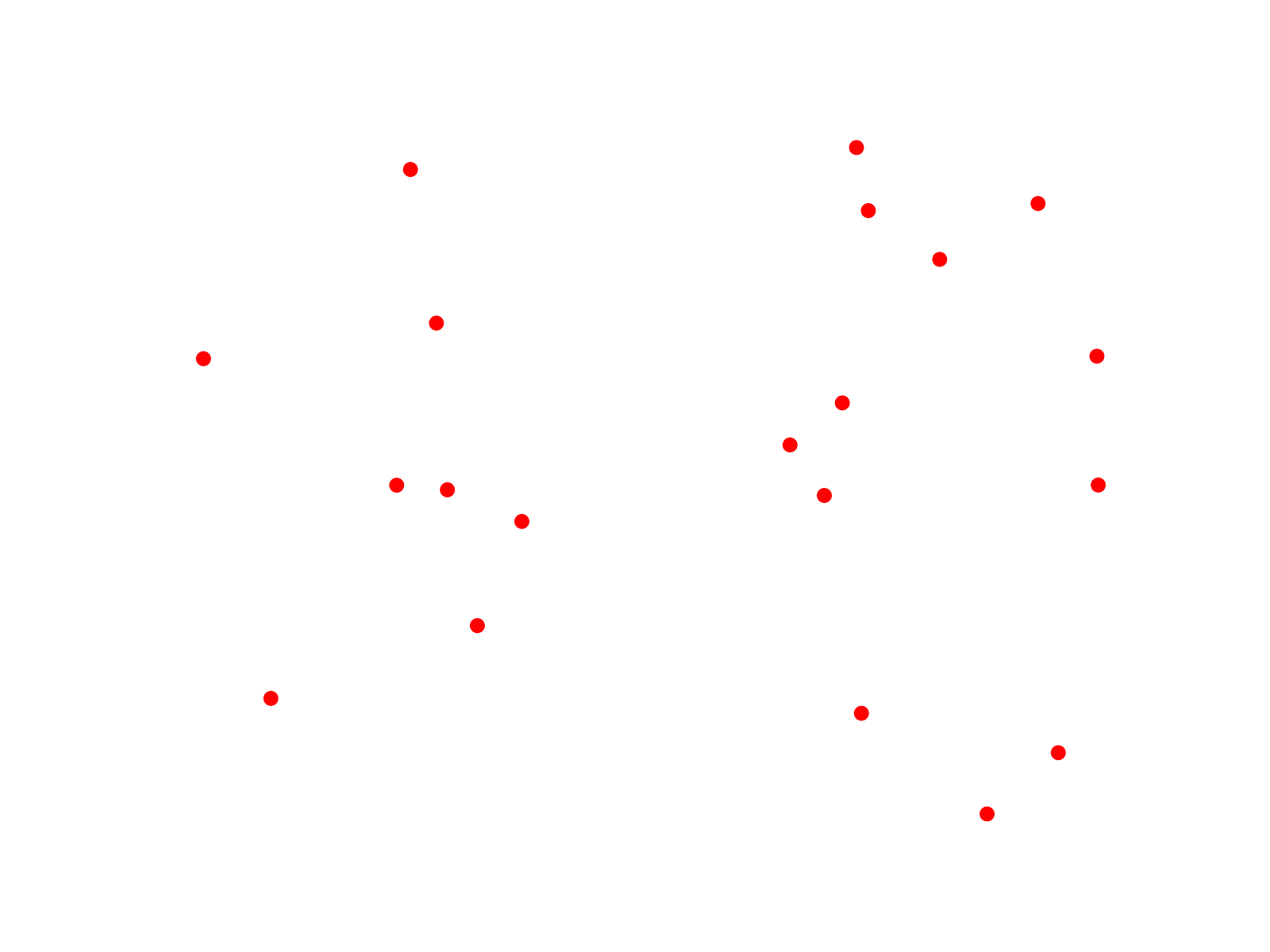}

    \vspace{.2cm}

    \includegraphics[trim={25mm 16mm 21mm 16mm},clip,width=.35\textwidth]{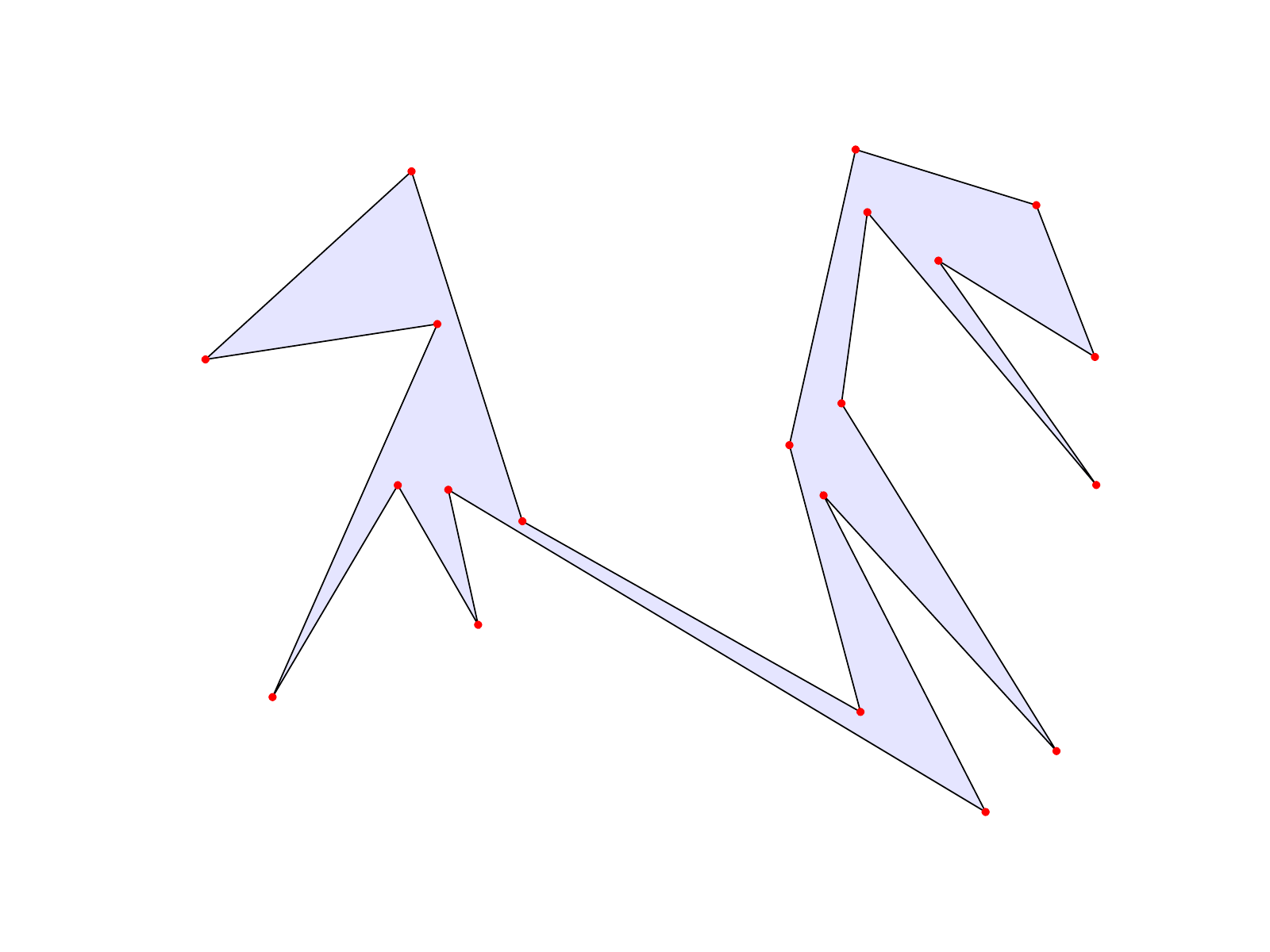}\hfil
    \includegraphics[trim={25mm 16mm 21mm 16mm},clip,width=.35\textwidth]{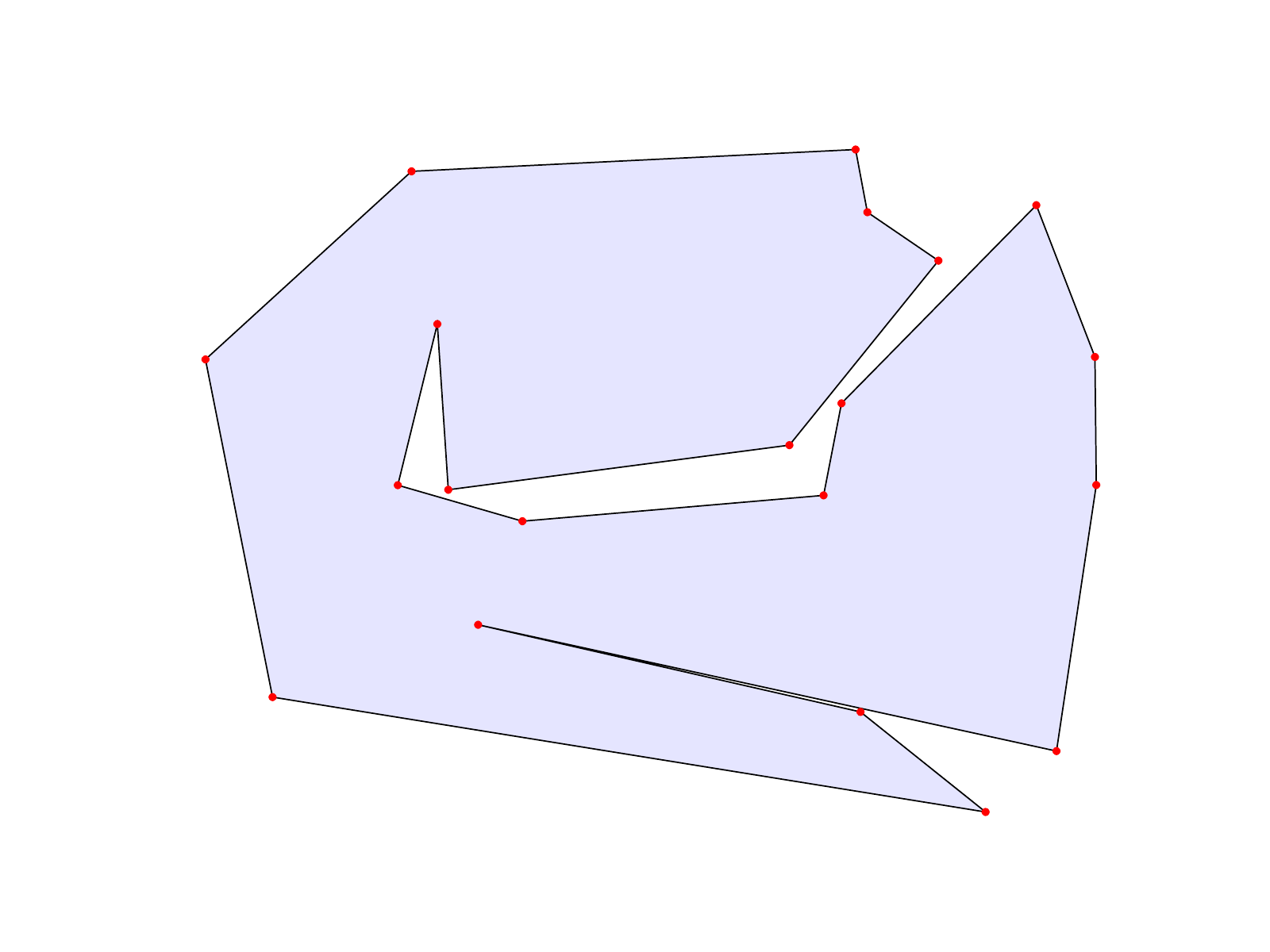}
  \caption{(Top) A set of 20 points. (Bottom Left) \textsc{Min-Area} solution. (Bottom Right) \textsc{Max-Area} solution.}
  \label{fig:example}
\end{figure}

While the classic geometric Traveling Salesman Problem (TSP) is to find a (simple)
polygon with a given set of vertices that has shortest perimeter, it is natural
to look for a simple polygon with a given set of vertices that minimizes
another basic geometric measure: the enclosed area.
The problem {\sc Min-Area} asks for a simple polygon with minimum enclosed
area, while {\sc Max-Area} demands one of maximum area; see Figure~\ref{fig:example}
for an illustration.

Both problem variants were shown to be $\mathcal{NP}$-complete by Fekete \cite{f-gtsp-92,fekete2000simple,fekete1993area},
who also showed that no {polynomial-time approximation scheme} (PTAS) exists for \textsc{Min-Area} problem and
gave a $\frac{1}{2}$-approximation algorithm for \textsc{Max-Area}.

\subsection{Related Work}
Most previous practical work has focused on finding heuristics for both problems.
Taranilla et al.~\cite{taranilla2011approaching} proposed
three different heuristics. 
Peethambaran~\cite{peethambaran2015randomized,peethambaran2016empirical} later proposed
randomized and greedy algorithms on solving \textsc{Min-Area} as well as the
$d$-dimensional variant of both \textsc{Min-Area} and \textsc{Max-Area}.
Considerable recent attention and progress was triggered by the 2019 CG Challenge,
which asked contestants to find good solutions for a wide spectrum of benchmark instances with up to
100,000 points; details are described in the survey by Demaine et al.~\cite{challenge19}.

Despite this focus, there has only been a limited amount of work on computing
provably optimal solutions for instances of interesting size. The only
notable exception is by Fekete et al.~\cite{fekete2015area}, who were able to solve
all instances of \textsc{Min-Area} with up to $n=14$ and some with up to $n=16$ points, as well as
all instances of \textsc{Max-Area} with up to $n=17$ and some with up to $n=19$ points.
This illustrates the inherent practical difficulty, which differs considerably from
the closely related TSP, for which even straightforward IP-based approaches can yield provably optimal solutions
for instances with hundreds of points, and sophisticated methods can solve instances with tens of thousands
of points.

\subsection{Our Results}
We present a systematic study of exact methods for \textsc{Min-Area} and \textsc{Max-Area} polygonizations.
We show that a number of careful enhancements can help to extend the range of instances
that can be solved to provable optimality, with different approaches working better
for the two problem variants. On the other hand, our work shows that the problems appear
to be practically harder than geometric optimization problems such as the Euclidean TSP.

	\section{Context and Preliminaries}
\label{sec:preliminaries}

A detailed description of background, context and further connections
can be found in the survey by Demaine et al.~\cite{challenge19}.

	\section{Tools}
\label{sec:tools}
We considered two models based on integer programming: an \emph{edge-based formulation}
(described in Section~\ref{sec:edge}) and a \emph{triangle-based formulation}
(described in Section~\ref{sec:triangle}). In addition, we developed a number
of further refinements and improvements (described in Section~\ref{sec:enhance}).

\subsection{Edge-Based Formulation}
\label{sec:edge}
\begin{figure}[h]
\centering
\begin{subfigure}[b]{.5\linewidth}
\centering
\resizebox{.6\columnwidth}{!}{\begin{tikzpicture}[line cap=round,line join=round,>=stealth,x=.4cm,y=.4cm]
\draw [<-] (2,0)-- (0,3);

\draw [->] (2,0)-- (3,3);

\draw [->] (3,3)-- (4,4);

\draw [->] (4,4)-- (1,5);

\draw [->] (1,5)-- (6,8);

\draw [->] (6,8)-- (0,10);

\draw [->] (0,10)-- (-6,6);

\draw [->] (-6,6)-- (-4,5);

\draw [->] (-4,5)-- (-6,1);

\draw [->] (-6,1)-- (0,3);
\begin{scriptsize}

\draw [dashed] (0,-3) -- (0,3);
\draw [dashed] (0,-3) -- (-6,1);
\draw [dashed] (0,-3) -- (2,0);
\draw [dashed] (0,-3) -- (3,3);
\draw [dashed] (0,-3) -- (4,4);
\draw [dashed] (0,-3) -- (1,5);
\draw [dashed] (0,-3) -- (6,8);
\draw [dashed] (0,-3) -- (0,10);
\draw [dashed] (0,-3) -- (-6,6);
\draw [dashed] (0,-3) -- (-4,5);

\draw [fill=red] (0,-3) circle (1.5pt) node[below] {$r$};
\end{scriptsize}
\end{tikzpicture}}%
\caption{Triangles of the polygon.}
\label{fig:reference_point_approach:polygon_triangles}
\end{subfigure}%
\begin{subfigure}[b]{.5\linewidth}
\centering
\resizebox{.6\columnwidth}{!}{\begin{tikzpicture}[line cap=round,line join=round,>=stealth,x=.4cm,y=.4cm]
\begin{scriptsize}

\draw [dashed] (0,-3) -- (0,3);
\draw [dashed] (0,-3) -- (-6,1);
\draw [dashed] (0,-3) -- (2,0);
\draw [dashed] (0,-3) -- (3,3);
\draw [dashed] (0,-3) -- (4,4);
\draw [dashed] (0,-3) -- (1,5);
\draw [dashed] (0,-3) -- (6,8);
\draw [dashed] (0,-3) -- (0,10);
\draw [dashed] (0,-3) -- (-6,6);
\draw [dashed] (0,-3) -- (-4,5);

\fill[fill=green, fill opacity=0.2] (0,-3)--(2,0)-- (3,3);
\fill[fill=green, fill opacity=0.2] (0,-3)--(-4,5)-- (-6,1);
\fill[fill=green, fill opacity=0.2] (0,-3)--(0,10)-- (6,8);
\fill[fill=green, fill opacity=0.2] (0,-3)--(-6,6)-- (0,10);
\fill[fill=green, fill opacity=0.2] (0,-3)--(4,4)-- (1,5);

\draw [<-, line width=.5pt] (2,0)-- (0,3);

\draw [->, line width=1pt] (2,0)-- (3,3);

\draw [->, line width=.5pt] (3,3)-- (4,4);

\draw [->, line width=1pt] (4,4)-- (1,5);

\draw [->, line width=.5pt] (1,5)-- (6,8);

\draw [->, line width=1pt] (6,8)-- (0,10);

\draw [->, line width=1pt] (0,10)-- (-6,6);

\draw [->, line width=.5pt] (-6,6)-- (-4,5);

\draw [->, line width=1pt] (-4,5)-- (-6,1);

\draw [->, line width=.5pt] (-6,1)-- (0,3);

\draw [fill=red] (0,-3) circle (1.5pt) node[below] {$r$};
\end{scriptsize}
\end{tikzpicture}}%
\caption{Positive edge triangles.}
\label{fig:reference_point_approach:polygon_positive}
\end{subfigure}%

\begin{subfigure}[b]{.5\textwidth}
\centering
\resizebox{.6\columnwidth}{!}{\begin{tikzpicture}[line cap=round,line join=round,>=stealth,x=.4cm,y=.4cm]

\begin{scriptsize}

\draw [dashed] (0,-3) -- (0,3);
\draw [dashed] (0,-3) -- (-6,1);
\draw [dashed] (0,-3) -- (2,0);
\draw [dashed] (0,-3) -- (3,3);
\draw [dashed] (0,-3) -- (4,4);
\draw [dashed] (0,-3) -- (1,5);
\draw [dashed] (0,-3) -- (6,8);
\draw [dashed] (0,-3) -- (0,10);
\draw [dashed] (0,-3) -- (-6,6);
\draw [dashed] (0,-3) -- (-4,5);

\fill[fill=red, fill opacity=0.2] (0,-3)--(-6,1)-- (0,3);
\fill[fill=red, fill opacity=0.2] (0,-3)--(2,0)-- (0,3);
\fill[fill=red, fill opacity=0.2] (0,-3)--(3,3)-- (4,4);
\fill[fill=red, fill opacity=0.2] (0,-3)--(1,5)-- (6,8);
\fill[fill=red, fill opacity=0.2] (0,-3)--(-6,6)-- (-4,5);

\draw [<-, line width=1pt] (2,0)-- (0,3);

\draw [->, line width=.5pt] (2,0)-- (3,3);

\draw [->, line width=1pt] (3,3)-- (4,4);

\draw [->, line width=.5pt] (4,4)-- (1,5);

\draw [->, line width=1pt] (1,5)-- (6,8);

\draw [->, line width=.5pt] (6,8)-- (0,10);

\draw [->, line width=.5pt] (0,10)-- (-6,6);

\draw [->, line width=1pt] (-6,6)-- (-4,5);

\draw [->, line width=.5pt] (-4,5)-- (-6,1);

\draw [->, line width=1pt] (-6,1)-- (0,3);

\draw [fill=red] (0,-3) circle (1.5pt) node[below] {$r$};
\end{scriptsize}
\end{tikzpicture}}%
\caption{Negative edge triangles.}
\label{fig:reference_point_approach:polygon_negative}
\end{subfigure}\begin{subfigure}[b]{.5\textwidth}
\centering
\resizebox{.6\columnwidth}{!}{\begin{tikzpicture}[line cap=round,line join=round,>=stealth,x=.4cm,y=.4cm]

\begin{scriptsize}

\draw [dashed] (0,-3) -- (0,3);
\draw [dashed] (0,-3) -- (-6,1);
\draw [dashed] (0,-3) -- (2,0);
\draw [dashed] (0,-3) -- (3,3);
\draw [dashed] (0,-3) -- (4,4);
\draw [dashed] (0,-3) -- (1,5);
\draw [dashed] (0,-3) -- (6,8);
\draw [dashed] (0,-3) -- (0,10);
\draw [dashed] (0,-3) -- (-6,6);
\draw [dashed] (0,-3) -- (-4,5);

\fill[fill=blue, fill opacity=0.2] (2,0)--(3,3)--(4,4)--(1,5)--(6,8)--(0,10)--(-6,6)--(-4,5)--(-6,1)--(0,3);

\draw [<-, line width=1pt] (2,0)-- (0,3);

\draw [->, line width=1pt] (2,0)-- (3,3);

\draw [->, line width=1pt] (3,3)-- (4,4);

\draw [->, line width=1pt] (4,4)-- (1,5);

\draw [->, line width=1pt] (1,5)-- (6,8);

\draw [->, line width=1pt] (6,8)-- (0,10);

\draw [->, line width=1pt] (0,10)-- (-6,6);

\draw [->, line width=1pt] (-6,6)-- (-4,5);

\draw [->, line width=1pt] (-4,5)-- (-6,1);

\draw [->, line width=1pt] (-6,1)-- (0,3);

\draw [fill=red] (0,-3) circle (1.5pt) node[below] {$r$};
\end{scriptsize}
\end{tikzpicture}}%
\caption{Calculated difference between (b) and (c)}
\label{fig:reference_point_approach:polygon_area}
\end{subfigure}
\caption[Area calculation in the boundary based approach]{Area calculation of a polygon using a reference point $r$ (red point)}
\label{fig:reference_point_approach:area_calculation}
\end{figure}
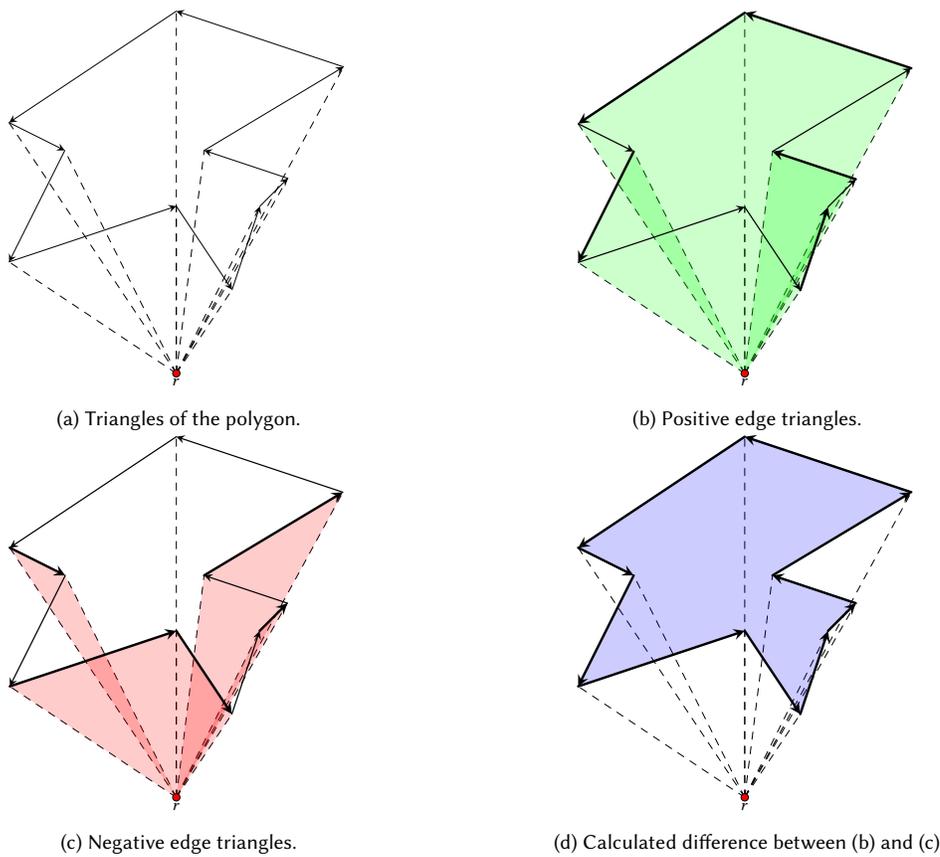

The first formulation is based on considering \emph{directed} edges of the polygon boundary.
We denote the boundary of a polygon $\Pol$ by $\partial {\Pol}$.
For two points $s_i, s_j \in S$, we consider the two directed half-edges
$e_{ij}$ and $e_{ji}$.  Let $E^r$ be the set of half-edges between the
points of $S$. As shown in Figure~\ref{fig:reference_point_approach:area_calculation},
the area $A_{\Pol}$ of a polygon $\Pol$ can be computed by adding the signed triangle areas $f_e$
that are formed by directed half-edges $e$ and an arbitrary, fixed reference point $r$:
$f_e$ is positive if $e$ and $r$ form a triangle for which $e$ has a counterclockwise orientation
along its boundary, and negative for a clockwise orientation. Therefore, we have to choose an appropriate
set $E_{\Pol}\subseteq E^r$ that optimizes the total area, as follows.

\begin{equation}
\label{eq:reference_point_approach:area_calculation}
A_{\Pol} = A(E_{\Pol}) = \sum_{e \in E_{\Pol}} f_e
\end{equation}


This gives rise to an integer program in which the choice of half-edges $e=(i,j)$ is modeled by 0-1 variables $z_e=z_{ij}$.

\begin{equation}
  \{\min , \max\} \sum_{e \in E^r} z_{e} \cdot f_e \label{ip:reference_point_approach:objective}\\
\end{equation}
\begin{align}
\vphantom{\sum_{i=1}^{m}} \forall s_i \in S: \qquad& \sum_{(j,i) \in \delta^+(s_i)} z_{ji} = 1 \label{ip:reference_point_approach:in_degree}\\
\vphantom{\sum_{i=1}^{m}} \forall s_i \in S: \qquad& \sum_{(i,j) \in \delta^-(s_i)} z_{ij} = 1 \label{ip:reference_point_approach:out_degree}\\
\vphantom{\sum_{i=1}^{m}} \forall e=\{i,j\} \in E: \qquad& z_{ij} + z_{ji} \leq 1 \label{ip:reference_point_approach:one_of_edges}\\
\vphantom{\sum_{i=1}^{m}} \forall \text{intersecting}\quad \{i,j\}, \{k,l\} \in E: \qquad& z_{ij} + z_{ji} + z_{kl} + z_{lk} \leq 1 \label{ip:reference_point_approach:intersections}\\
\vphantom{\sum_{i=1}^{m}} (\forall \text{slabs}\quad L) (\forall m=1,\ldots,\vert L \vert): \qquad& \sum_{i=1}^{m} z_{e_{i_L}^{lr}}-z_{e_{i_L}^{rl}} \begin{array}{l}
	\leq 1\\
	\geq 0
	\end{array} \label{ip:reference_point_approach:slabs}\\
\vphantom{\sum_{i=1}^{m}} \forall \emptyset \neq D \subsetneq S: \qquad& \begin{array}{l}\sum_{(k,l) \in \delta^-(D)} z_{kl} \geq 1 \\ \sum_{(k,l) \in \delta^+(D)} z_{kl} \geq 1\end{array}  \label{ip:reference_point_approach:subtours}\\
\forall {i,j} \in E: \qquad& z_{ij},z_{ji} \in \{0,1\}
\end{align}

The \emph{objective function} \eqref{ip:reference_point_approach:objective} arises from signed triangle areas, as described.
The constraints \eqref{ip:reference_point_approach:in_degree} and
\eqref{ip:reference_point_approach:out_degree} ensure that each point $s_i \in
S$ has one outgoing edge and one incoming edge in the resulting polygon.
Furthermore, constraints \eqref{ip:reference_point_approach:one_of_edges}
guarantee, that for each possible edge, only one of the half-edges can be in
$\partial P$. Intersecting edges in the resulting polygon are excluded by
constraints~\eqref{ip:reference_point_approach:intersections}.

\begin{figure}[h]
\centering
\resizebox{5.5cm}{!}{\begin{tikzpicture}[line cap=round,line join=round,x=1cm,y=1cm]
\draw [line width=1pt] (-2,0)-- (2,1);
\draw [line width=1pt] (-2,0)-- (1,3);

\draw [line width=1pt] (-4,1)-- (2,1);
\draw [line width=1pt] (-4,1)-- (1,3);

\draw [line width=1pt] (-3,4)-- (2,1);
\draw [line width=1pt] (-3,4)-- (1,3);

\draw [line width=1pt] (2,1) -- (1,3);
\draw [line width=1pt] (-4,1)-- (-3,4);
\draw [line width=1pt] (-2,0)-- (-3,4);
\draw [line width=1pt] (-4,1)-- (-2,0);

\draw [dashed, line width=1pt] (-3,4.5)-- (-3,-1);
\draw [dashed, line width=1pt] (-4,4.5)-- (-4,-1);
\draw [dashed, line width=1pt] (-2,4.5)-- (-2,-1);
\draw [dashed, line width=1pt] (1,4.5)-- (1,-1);
\draw [dashed, line width=1pt] (2,4.5)-- (2,-1);

\draw[stealth-stealth, line width=1.5pt] (-2,-1.5) -- node[midway,fill=white] {slab}  (1,-1.5);

\draw[|-latex, line width=1.5pt, blue] (-0.5,-0.5) -- (-0.5,4);

\end{tikzpicture}}%
\caption{Visualization of slabs in a polygon}
\label{fig:reference_point_approach:slabs}
\end{figure}
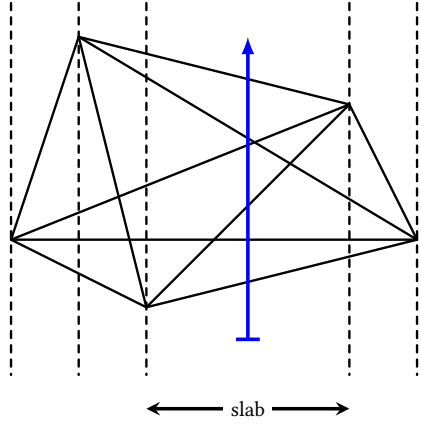

\begin{figure}[h]
\begin{subfigure}[b]{.5\textwidth}
\centering
  \resizebox{2.5cm}{!}{\begin{tikzpicture}[line cap=round,line join=round,x=1cm,y=1cm]

\draw [dashed, line width=.5pt] (-2,0)-- (1,0.75);
\draw [dashed, line width=.5pt] (-2,0)-- (1,3);

\draw [dashed, line width=.5pt] (-2,1)-- (1,1);
\draw [dashed, line width=.5pt] (-2,1.8)-- (1,3);

\draw [dashed, line width=.5pt] (-2,3.4)-- (1,1.6);
\draw [dashed, line width=.5pt] (-2,3.75)-- (1,3);

\draw [line width=1pt] (-2,4.5)-- (-2,-1);
\draw [line width=1pt] (1,4.5)-- (1,-1);

\draw [-stealth, line width=1pt] (-2,0)-- (1,0.75);
\draw [stealth-, line width=1pt] (-2,1)-- (1,1);
\draw [-stealth, line width=1pt] (-2,1.8)-- (1,3);
\draw [stealth-, line width=1pt] (-2,3.75)-- (1,3);

\draw[|-latex, line width=1.5pt, blue] (-0.5,-0.5) -- (-0.5,4);

\end{tikzpicture}}
\end{subfigure}\begin{subfigure}[b]{.5\textwidth}
\centering
  \resizebox{2.5cm}{!}{\begin{tikzpicture}[line cap=round,line join=round,x=1cm,y=1cm]

\draw [dashed, line width=.5pt] (-2,0)-- (1,0.75);
\draw [dashed, line width=.5pt] (-2,0)-- (1,3);

\draw [dashed, line width=.5pt] (-2,1)-- (1,1);
\draw [dashed, line width=.5pt] (-2,1.8)-- (1,3);

\draw [dashed, line width=.5pt] (-2,3.4)-- (1,1.6);
\draw [dashed, line width=.5pt] (-2,3.75)-- (1,3);

\draw [line width=1pt] (-2,4.5)-- (-2,-1);
\draw [line width=1pt] (1,4.5)-- (1,-1);

\draw [-stealth, line width=1pt] (-2,0)-- (1,0.75);
\draw [stealth-, line width=1pt] (-2,1)-- (1,1);
\draw [stealth-, line width=1pt, red] (-2,3.4)-- (1,1.6);
\draw [-stealth, line width=1pt] (-2,3.75)-- (1,3);

\draw[|-latex, line width=1.5pt, blue] (-0.5,-0.5) -- (-0.5,4);

\end{tikzpicture}}
\end{subfigure}

\caption[Visualization of slab constraints]{Visualization of slab constraints. Left side shows a valid configuration. The right side shows a violated slab constraint.}
\label{fig:reference_point_approach:slabs_allowed}
\end{figure}
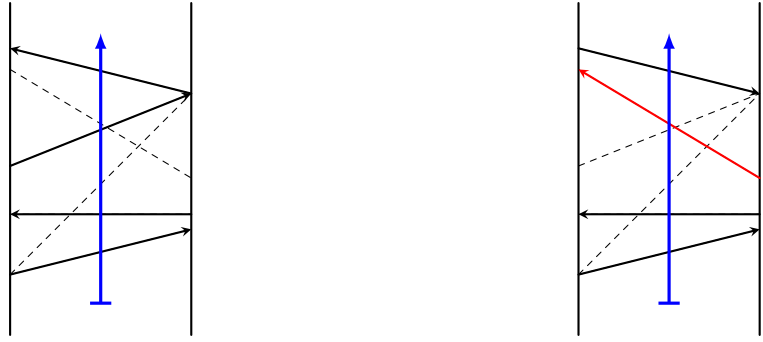

The next set of inequalities \eqref{ip:reference_point_approach:slabs} are called \emph{slab constraints};
they ensure that the polygon is oriented in a counterclockwise manner.
A \emph{slab} $L$ is a vertical strip bounded by the $x$-coordinates of two consecutive points in
the order of $x$-coordinates of points. Figure~\ref{fig:reference_point_approach:slabs} shows the slabs of a given
point set. The edges of slab $L$ get ordered by the $y$-coordinate at point
$c_L=\frac{c^{x_{max}}_L-c^{x_{min}}_L}{2}$, where $c^{x_{max}}_L$ ($c^{x_{min}}_L$, resp.) is the $x$-coordinate of the right (left, resp.) boundary of slab $L$.
Figure~\ref{fig:reference_point_approach:slabs} illustrates $c_L$ with the
blue arrow within the indicated slab. Now the bottommost chosen edge has
to be oriented from left to right and the topmost one from right
to left, while chosen edges in between have to alternate in their
direction. This is enforced by the slab constraints
for all possible sums $\sum_{i=1}^{m} z_{e_{i_L}^{lr}}-z_{e_{i_L}^{rl}}$ with
$m=1,\dots,\vert L \vert$, where $e_{i_L}^{lr}$ ($e_{i_L}^{rl}$, resp.) denotes the $i$-th bottom most half-edge along slab $L$ going from left to right (from right to left, resp.).
Figure~\ref{fig:reference_point_approach:slabs_allowed} shows two possible
configurations of a slab. The left one is a valid slab and satisfies all
inequalities \eqref{ip:reference_point_approach:slabs}, while the right one does
violate the constraint
$$ \underbrace{e^{lr}_1-e_1^{rl}}_{=1} + \underbrace{e^{lr}_2-e_2^{rl}}_{=-1} + \underbrace{e^{lr}_3-e_3^{rl}}_{=0} + \underbrace{e^{lr}_4-e_4^{rl}}_{=0} +
\underbrace{e^{lr}_5-e_5^{rl}}_{=-1} = -1 \ngeq 0 \quad m=5$$
Note that we have to add inequalities for all $m = 1,\dotsc,\vert L \vert$. For the previous example all other inequalities are satisfied.
\begin{align*}
e^{lr}_1-e_1^{rl} &&&&&&&&&&&=1 \begin{array}{l}\geq 0\\\leq 1\end{array} \quad m=1\\
e^{lr}_1-e_1^{rl} &+
&e^{lr}_2-e_2^{rl} &&&&&&&&&= 0\begin{array}{l}\geq 0\\\leq 1\end{array} \quad m=2\\
e^{lr}_1-e_1^{rl} &+
&e^{lr}_2-e_2^{rl} &+
&e^{lr}_3-e_3^{rl}
&&&&&&&= 0\begin{array}{l}\geq 0\\\leq 1\end{array} \quad m=3\\
e^{lr}_1-e_1^{rl} &+
&e^{lr}_2-e_2^{rl} &+
&e^{lr}_3-e_3^{rl} &+
&e^{lr}_4-e_4^{rl}
&&&&&= 0\begin{array}{l}\geq 0\\\leq 1\end{array} \quad m=4\\
\underbrace{e^{lr}_1-e_1^{rl}}_{=1} &+ &\underbrace{e^{lr}_2-e_2^{rl}}_{=-1} &+ &\underbrace{e^{lr}_3-e_3^{rl}}_{=0} &+ &\underbrace{e^{lr}_4-e_4^{rl}}_{=0} &+
&\underbrace{e^{lr}_5-e_5^{rl}}_{=-1} &+
&\underbrace{e^{lr}_6-e_6^{rl}}_{=1} &= 0\begin{array}{l}\geq 0\\\leq 1\end{array} \quad m=6
\end{align*}

The last set of constraints \eqref{ip:reference_point_approach:subtours} are
\emph{subtour constraints}; they ensure that all non-trivial subsets of vertices
have at least one incoming and one outgoing edge.
These constraints are very common for many related
optimization problems, including (in undirected form) for the TSP.

Overall, the size of the resulting IP is as follows.

\begin{itemize}
  \item There is a total of $O(2n) = O(n)$  point degree constraints \eqref{ip:reference_point_approach:in_degree} and \eqref{ip:reference_point_approach:out_degree}, two for each vertex.
  \item There is a total of $O(\binom{n}{2})=O(n^2)$ half-edge constraints, one for each half-edge.
  \item There is a total of $O(n^4)$ intersections constraints \eqref{ip:reference_point_approach:intersections}, one for each pair of intersecting edges.
  \item There is a total of $O(n^3)$ slab constraints \eqref{ip:reference_point_approach:slabs}, one for each combination of one of the $n-1$ slabs and the $O(n^2)$ possible edges crossing it.
  \item There is a total of $O(2^n)$ subtour constraints \eqref{ip:reference_point_approach:subtours}, one for each non-trivial subset of vertices.
\end{itemize}

In the practical implementation we cannot add all subtour constraints and try
to avoid adding intersection constraints before starting the branch and cut
algorithm. While solving the IP we get access to partial solutions and only add
new constraints when necessary. Therefore, we start with a slim IP with only $O(n +
n^2 + n^3) = O(n^3)$ constraints and $O(n^2)$ variables. During the branch-and-cut
algorithm at most $O(2^n)$ constraints are added.

\subsection{Triangle-Based Formulation}
\label{sec:triangle}

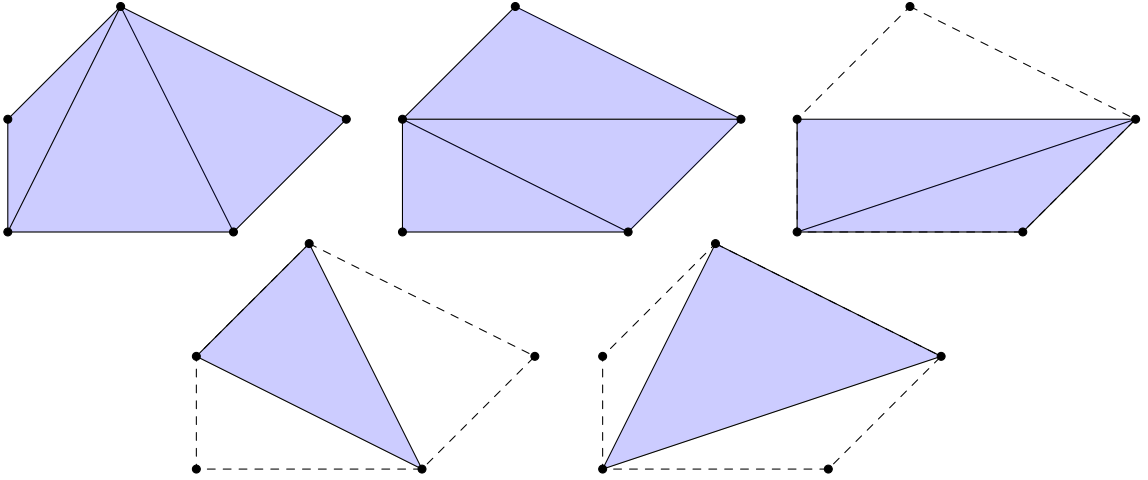
\begin{figure}[t]
\centering
  \begin{tikzpicture}[line cap=round,line join=round,x=1.5cm,y=1.5cm]
\coordinate (A) at (0,1);
\coordinate (B) at (2,0);
\coordinate (C) at (1,-1);
\coordinate (D) at (-1,-1);
\coordinate (E) at (-1,0);

\draw [] (A) -- (B) -- (C) -- (D) -- (E) -- (A);


\draw[fill=blue, opacity=0.2] (A) -- (E) -- (D);
\draw[] (A) -- (D);

\draw[fill=blue, opacity=0.2] (A) -- (D) -- (C);
\draw[] (A) -- (C);

\draw[fill=blue, opacity=0.2] (A) -- (B) -- (C);

\begin{scriptsize}
\draw [fill=black] (A) circle (1.5pt);
\draw [fill=black] (B) circle (1.5pt);
\draw [fill=black] (C) circle (1.5pt);
\draw [fill=black] (D) circle (1.5pt);
\draw [fill=black] (E) circle (1.5pt);

\end{scriptsize}
\end{tikzpicture}\hfill
\begin{tikzpicture}[line cap=round,line join=round,x=1.5cm,y=1.5cm]
\coordinate (A) at (0,1);
\coordinate (B) at (2,0);
\coordinate (C) at (1,-1);
\coordinate (D) at (-1,-1);
\coordinate (E) at (-1,0);

\draw [] (A) -- (B) -- (C) -- (D) -- (E) -- (A);


\draw[fill=blue, opacity=0.2] (A) -- (E) -- (B);
\draw[] (E) -- (B);

\draw[fill=blue, opacity=0.2] (B) -- (E) -- (C);
\draw[] (E) -- (C);

\draw[fill=blue, opacity=0.2] (C) -- (D) -- (E);

\begin{scriptsize}
\draw [fill=black] (A) circle (1.5pt);
\draw [fill=black] (B) circle (1.5pt);
\draw [fill=black] (C) circle (1.5pt);
\draw [fill=black] (D) circle (1.5pt);
\draw [fill=black] (E) circle (1.5pt);

\end{scriptsize}
\end{tikzpicture}\hfill
\begin{tikzpicture}[line cap=round,line join=round,x=1.5cm,y=1.5cm]
\coordinate (A) at (0,1);
\coordinate (B) at (2,0);
\coordinate (C) at (1,-1);
\coordinate (D) at (-1,-1);
\coordinate (E) at (-1,0);

\draw [dashed] (A) -- (B) -- (C) -- (D) -- (E) -- (A);


\draw[fill=blue, opacity=0.2] (E) -- (D) -- (B);
\draw[] (E) -- (B);
\draw[] (D) -- (B);
\draw[] (D) -- (E);

\draw[fill=blue, opacity=0.2] (B) -- (C) -- (D);
\draw[] (B) -- (C);
\draw[] (D) -- (C);

\begin{scriptsize}
\draw [fill=black] (A) circle (1.5pt);
\draw [fill=black] (B) circle (1.5pt);
\draw [fill=black] (C) circle (1.5pt);
\draw [fill=black] (D) circle (1.5pt);
\draw [fill=black] (E) circle (1.5pt);

\end{scriptsize}
\end{tikzpicture}\\
\begin{tikzpicture}[line cap=round,line join=round,x=1.5cm,y=1.5cm]
\coordinate (A) at (0,1);
\coordinate (B) at (2,0);
\coordinate (C) at (1,-1);
\coordinate (D) at (-1,-1);
\coordinate (E) at (-1,0);

\draw [dashed] (A) -- (B) -- (C) -- (D) -- (E) -- (A);


\draw[fill=blue, opacity=0.2] (A) -- (E) -- (C);
\draw[] (E) -- (C);
\draw[] (A) -- (C);
\draw[] (A) -- (E);

\begin{scriptsize}
\draw [fill=black] (A) circle (1.5pt);
\draw [fill=black] (B) circle (1.5pt);
\draw [fill=black] (C) circle (1.5pt);
\draw [fill=black] (D) circle (1.5pt);
\draw [fill=black] (E) circle (1.5pt);

\end{scriptsize}
\end{tikzpicture}\hspace{20pt}
\begin{tikzpicture}[line cap=round,line join=round,x=1.5cm,y=1.5cm]
\coordinate (A) at (0,1);
\coordinate (B) at (2,0);
\coordinate (C) at (1,-1);
\coordinate (D) at (-1,-1);
\coordinate (E) at (-1,0);

\draw [dashed] (A) -- (B) -- (C) -- (D) -- (E) -- (A);


\draw[fill=blue, opacity=0.2] (A) -- (B) -- (D);
\draw[] (D) -- (B);
\draw[] (A) -- (D);
\draw[] (A) -- (B);

\begin{scriptsize}
\draw [fill=black] (A) circle (1.5pt);
\draw [fill=black] (B) circle (1.5pt);
\draw [fill=black] (C) circle (1.5pt);
\draw [fill=black] (D) circle (1.5pt);
\draw [fill=black] (E) circle (1.5pt);

\end{scriptsize}
\end{tikzpicture}
  \caption[Empty triangle amount bound]{A set of five points and its ten empty triangles.}
  \label{fig:triangulation_approach:max_triangles}
\end{figure}

An alternative is the {triangle-based formulation}, which considers the set $T(P)$ of
possibly $\binom{n}{3}$ many empty triangles of a point set $P$; see Figure~\ref{fig:triangulation_approach:max_triangles}
for an illustration.
Making use of the fact that a simple polygon with $n$ vertices consists of $(n-2)$ empty
triangles with non-intersection interiors, we get the following IP formulation, in which the presence of
an empty triangle $\triangle$ with unsigned area $f_\triangle$ is described by a 0-1 variable $x_\subscripttriangle$.

\begin{equation}
  \{\min , \max\} \sum_{\subscripttriangle \in T(P)} f_\subscripttriangle \cdot x_\subscripttriangle \label{ip:triangulation_approach:objective}\\
\end{equation}
\begin{align}
\vphantom{\sum_{i=1}^{n}} \qquad& \sum_{\subscripttriangle \in T} x_\subscripttriangle = n-2 \label{ip:triangulation_approach:triangle_count}\\
\vphantom{\sum_{i=1}^{n}} \forall s_i \in S: \qquad& \sum_{\subscripttriangle \in \delta(s_i)} x_\subscripttriangle \geq 1 \label{ip:triangulation_approach:in_each_point}\\
\vphantom{\sum_{i=1}^{m}} \forall \text{intersecting}\quad \triangle_i, \triangle_j \in T(P): \qquad& x_{\subscripttriangle_i} + x_{\subscripttriangle_j} \leq 1 \label{ip:triangulation_approach:intersections}\\
\vphantom{\sum_{i=1}^{n}} \forall \emptyset \neq D \subsetneq T(P), \vert D \vert \leq n-3: \qquad& \sum_{\subscripttriangle \in D} x_\subscripttriangle - \sum_{\subscripttriangle \in \delta(D)} x_\subscripttriangle\leq  \vert D \vert - 1 \label{ip:triangulation_approach:subtours}\\
\vphantom{\sum_{i=1}^{n}} \forall \triangle \in T(P): \qquad& x_\subscripttriangle \in \{0,1\}
\end{align}

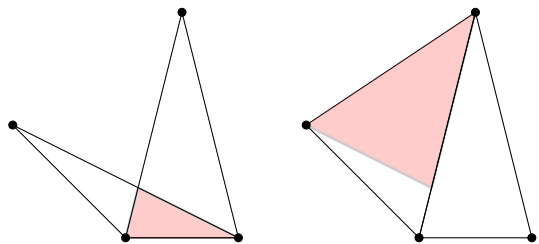
\begin{figure}[h!]
\centering
  \begin{tikzpicture}[line cap=round,line join=round,x=1.5cm,y=1.5cm]
\coordinate (A) at (0,0);
\coordinate (B) at (1,0);
\coordinate (C) at (0.5,2);
\coordinate (D) at (-1,1);
\coordinate (E) at (2/18,8/18);

\draw [] (A) -- (B) -- (C) -- cycle;
\draw [] (A) -- (B) -- (D) -- cycle;


\draw[fill=red, opacity=0.2, line width=1pt] (A) -- (E) -- (B);

\begin{scriptsize}
\draw [fill=black] (A) circle (1.5pt);
\draw [fill=black] (B) circle (1.5pt);
\draw [fill=black] (C) circle (1.5pt);
\draw [fill=black] (D) circle (1.5pt);

\end{scriptsize}
\end{tikzpicture}\hspace{20pt}
\begin{tikzpicture}[line cap=round,line join=round,x=1.5cm,y=1.5cm]
\coordinate (A) at (0,0);
\coordinate (B) at (1,0);
\coordinate (C) at (0.5,2);
\coordinate (D) at (-1,1);
\coordinate (E) at (2/18,8/18);

\draw [] (A) -- (B) -- (C) -- cycle;
\draw [] (A) -- (C) -- (D) -- cycle;


\draw[fill=red, opacity=0.2, line width=1pt] (D) -- (E) -- (C);

\begin{scriptsize}
\draw [fill=black] (A) circle (1.5pt);
\draw [fill=black] (B) circle (1.5pt);
\draw [fill=black] (C) circle (1.5pt);
\draw [fill=black] (D) circle (1.5pt);

\end{scriptsize}
\end{tikzpicture}
  \caption[Triangulations of a point set]{Triangulations of a point set with (right) and without (left) intersection constraints. The difference of both areas is the area difference of both red triangles.}
  \label{fig:triangulation_approach:intersectons}
\end{figure}

The objective function \eqref{ip:triangulation_approach:objective} is the sum
over the chosen triangles areas.
Triangle constraint
\eqref{ip:triangulation_approach:triangle_count} ensures that we chose exactly $n-2$
triangles, which is the number of triangles in a triangulation of a simple polygon.
Furthermore, point constraints \eqref{ip:triangulation_approach:in_each_point} guarantee that a solution has
at least one adjacent triangle at each point $s_i \in S$.
Finally, intersection constraints \eqref{ip:triangulation_approach:intersections} ensure that we only
select triangles with disjoint interiors. As shown in Figure~\ref{fig:triangulation_approach:intersectons},
these are indeed necessary, even when minimizing total area.

\begin{figure}
	\centering
	\includegraphics[scale=.6]{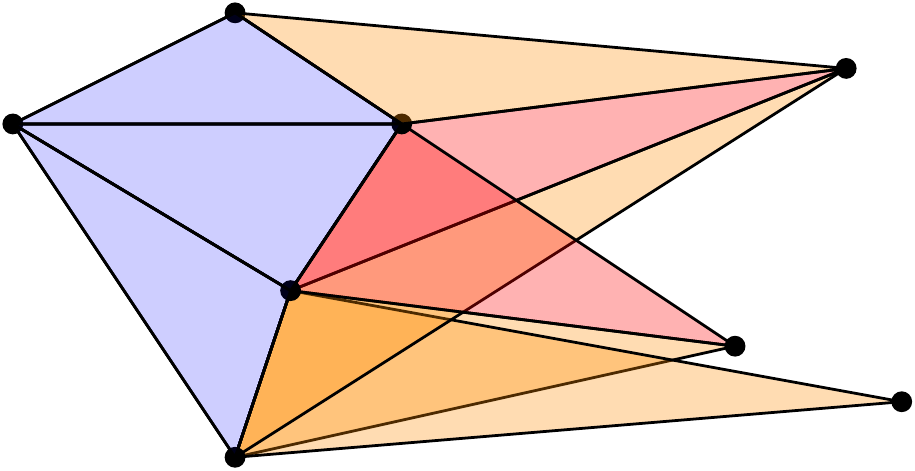}
	\caption{Visualization of the subtour constraint.
	Blue triangles represent the set $D$. 
	To satisfy the subtour constraint, either one of the red or orange triangles must be added, or one of the blue triangles must be removed.}
	\label{fig:subtour_constraint}
\end{figure}

Finally, the subtour constraints  \eqref{ip:triangulation_approach:subtours} ensure that the set of selected triangles
forms a simple polygon:
Either all triangles of $D$ are part of the solution and at least one new triangle must be adjacent to the boundary of $D$ (i.e., a triangle of $\delta(D)$), or one triangle of $D$ is not part of the solution (see Fig.~\ref{fig:subtour_constraint} for a visualization).

Overall, the size of the resulting IP is as follows.

\begin{itemize}
  \item There is a single triangle constraint of type \eqref{ip:triangulation_approach:triangle_count}.
  \item There are $O(n)$ point constraints \eqref{ip:triangulation_approach:in_each_point}, one for each point $s_i \in S$.
  \item There are $O(n^6)$ intersection constraints \eqref{ip:triangulation_approach:intersections}, one for each intersecting pair of the $O(n^3)$ triangles.
  \item There are $O(2^{n^3})$ subtour constraints \eqref{ip:triangulation_approach:subtours}, one for each $D\subsetneq T$ with $0 < \vert D \vert \leq n-3$ and all triangles in $D$.
\end{itemize}

Because of the enormous number of subtour constraints and the fact that most of the intersection constraints are not needed,
both types are dynamically added during the branch-and-cut algorithm.

\subsection{Enhancing the Integer Programs}
\label{sec:enhance}

Given the considerable size of the described IP formulations, we employed a number of enhancements to improve efficiency.

\subsubsection{Convex Hull}
\label{sec:convex_hull}

The area of the convex hull is an upper bound for every polygonization of a given point set.
Its combinatorial structure allows omitting a number of constraints:
The edge $e$ between two non-adjacent points on the boundary of the convex hull
divides the point set into two separate pieces, 
so we can remove edges between two non-adjacent points of the convex hull from our set of variables.


\subsubsection{Initial Solutions}
When solving an optimization problem, it helps to start with an initial
integer solution, which leads to a polygon $\Pol_{initial}$. Until no better
solution has been found, the area of $\Pol_{initial}$ helps the bounding process
to cut off subtrees of possible solutions. 
Starting with a solution that is very close to the optimal solution can accelerate the
computation speed a lot. As described in the survey article~\cite{challenge19}, there is an
approximation method by Fekete~\cite{f-gtsp-92} for the \textsc{Max-Area} problem, which guarantees to be at
most $\frac{1}{2}$ as large as the optimal solution. 
For the minimization problem 
we use a heuristic without performance guarantee.

We also used a simple greedy approach to obtain an initial value for \textsc{Min-Area}, based
on a heuristic of Taranilla et al.~\cite{taranilla2011approaching}.
The algorithm modifies an existing polygon $\Pol$
until all points are on the boundary.
\begin{enumerate}
  \item Compute the convex hull of the point set $S$, resulting in the initial polygon $\Pol$.
  \item Among all points inside $\Pol$ that are not part of $\Pol$, chose one point $s_1$. The point forms an empty triangle with some edge $\{s_2,s_3\}$ of $\Pol$, does not intersect with $\partial \Pol$ and is maximum in size. If no point $s_1$ inside $\Pol$ exists, we are finished, because $\Pol$ is a simple polygon that has all points of $S$ on its boundary.
  \item Remove edge $\{s_2,s_3\}$ from $\Pol$ and add edges $\{s_1,s_2\}, \{s_1,s_3\}$. Repeat Step 2.
\end{enumerate}

\begin{figure}[h!]
  \centering
  \hfill
\begin{tikzpicture}[line cap=round,line join=round,x=1cm,y=1cm]

\fill (0,0) circle (2pt);
\fill (2.3,2) circle (2pt);
\fill (0.6,4) circle (2pt);
\fill (0,5) circle (2pt);
\fill (-2,3) circle (2pt);
\fill (-3,-1) circle (2pt);

\fill (0,1.5) circle (2pt);
\fill (0.4,1.9) circle (2pt);
\fill (-1.3,3) circle (2pt);

\draw [] (0,0) -- (2.3,2) -- (0.6,4) -- (0,5) -- (-2,3) -- (-3, -1) -- cycle;
\draw [dashed] (-2,3) node[left] {$s_2$} -- (0,1.5) node[right] {$s_1$} -- (-3, -1) node[left] {$s_3$};
\end{tikzpicture}
\hfill
\begin{tikzpicture}[line cap=round,line join=round,x=1cm,y=1cm]

\fill (0,0) circle (2pt);
\fill (2.3,2) circle (2pt);
\fill (0.6,4) circle (2pt);
\fill (0,5) circle (2pt);
\fill (-2,3) circle (2pt);
\fill (-3,-1) circle (2pt);

\fill (0,1.5) circle (2pt);
\fill (0.4,1.9) circle (2pt);
\fill (-1.3,3) circle (2pt);

\draw [] (0,0) -- (2.3,2) -- (0.6,4) -- (0,5) -- (-2,3) node[left] {$s_2$} -- (0,1.5) node[right] {$s_1$} -- (-3, -1)  node[left] {$s_3$} -- cycle;

\end{tikzpicture}
\hfill
  \caption[The \textsc{Greedy Min-Area} algorithm]{A possible step two of the \textsc{Greedy Min-Area} algorithm}
  \label{fig:greedy_map}
\end{figure}
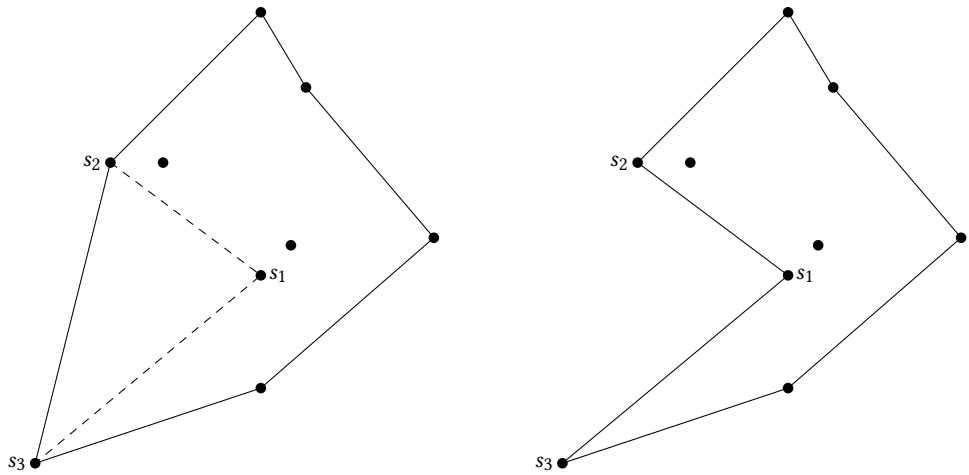

Figure~\ref{fig:greedy_map} illustrates a possible second step of the
\textsc{Greedy Min-Area} algorithm. The complexity of the first step is $O(n \log
n)$, because we need to compute a convex hull. For the second step we need to
consider each of the $n$ edges of $\Pol$ and compute triangles with possibly
$n$ points inside $\Pol$. We need to check whether the triangle is empty and
whether the triangle does not cross any of the $n$ edges of $\Pol$. The second
step has to be carried out for every point that is not part of the convex hull, i.e.,
$n$ times. This leads to an overall complexity of $O(n^4)$. The third step can
be done in constant time $O(1)$. Overall, this leads to a complexity of
of $O(n\log n + n^4 + 1) = O(n^4)$  for \textsc{Greedy Min-Area}. If we take away triangles with
minimum in step two of the algorithm, we get an heuristic for the maximization problem.
We call this variant \textsc{Greedy Max-Area}.

\subsubsection{Intersections}

\paragraph{Intersection Cliques}
If any pair $o_i,o_j$ in a given set $C=\{o_1,\ldots,o_k\}$ of objects (which may be edges or triangles)
intersect, they form an \emph{intersection clique}. Clearly, this allows replacing the $\Theta(k^2)$
pairwise intersection constraints by a single one, as follows.
$$\sum_{i=1}^k x_i \leq 1.$$
Because of the NP-completeness of finding maximum cardinality cliques, we simply use
\emph{maximal cliques}.

We only add intersection constraints incrementally, i.e., whenever we get a new integer solution,
we add new violated intersection constraints.
Because it is very time-consuming to compute all maximal
cliques in every iteration, we add cliques by consecutively adding
more edges to an existing clique $C$ until no edge can be found, which
intersects all edges in $C$. We start with $C=\{o_1,o_2\}$ with $o_1,o_2$ being
two intersecting objects of the solution. We then try to add more objects of the
current solution to the clique until no such object can be found. Afterwards all
other objects are considered, until no object
can be added to $C$. This provides good solutions in practice.

\begin{figure}[h!]
  \centering
  \includegraphics[scale=.7]{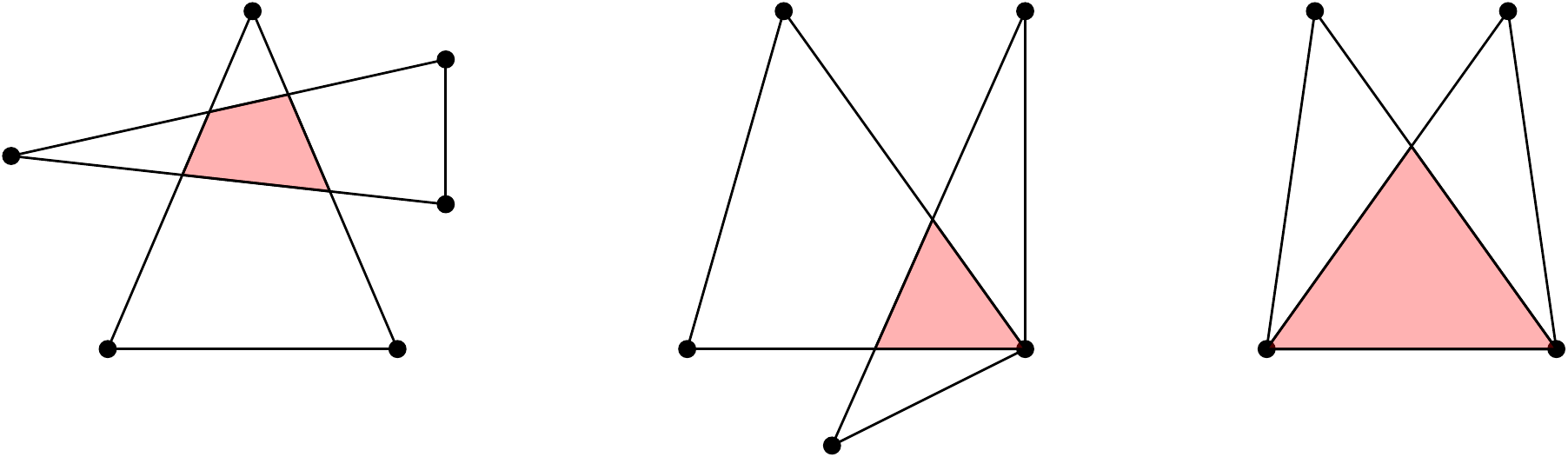}
  \caption{All three types of possible intersections in the triangle-based approach.}
  \label{fig:intersections_triangulation_example}
\end{figure}

\paragraph{Halfspace Constraints}
\label{sec:halfspace_constraints}
For the triangle-based approach,
we introduce the concept of \emph{halfspace constraints}.
The goal is to reduce the number of intersections that need to be added during the optimization process by
excluding obvious intersections from the beginning.
Figure~\ref{fig:intersections_triangulation_example} shows the two types of
intersections which may occur. Two intersecting triangles may share no point, a single point, or two points.

\begin{figure}[h!]
   \centering
   \includegraphics[scale=0.7]{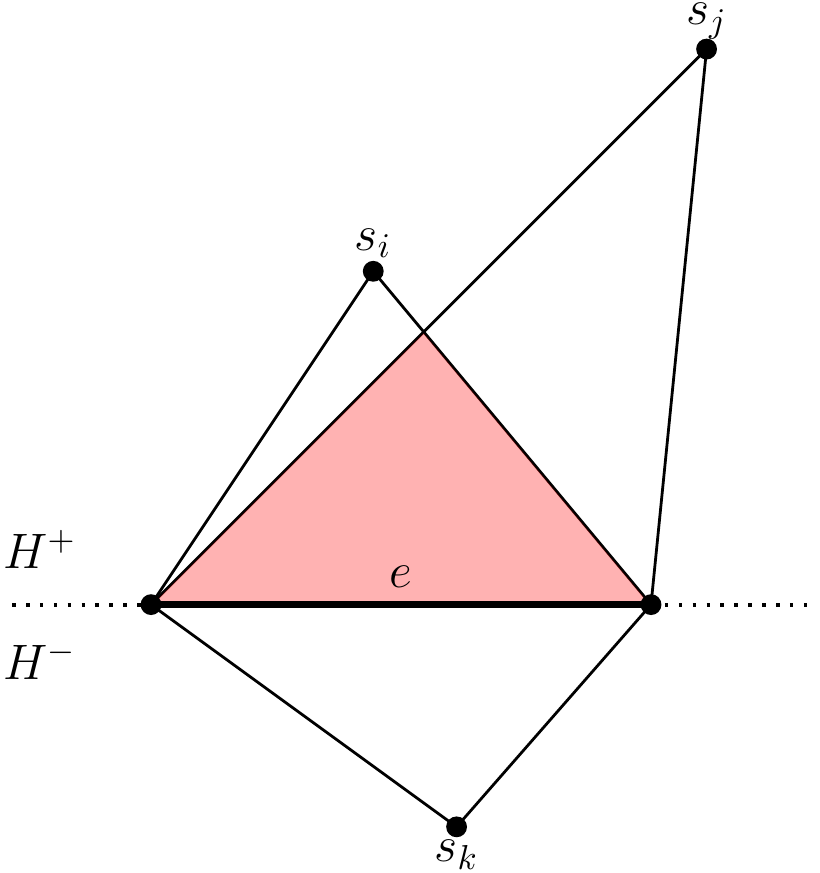}
   \caption{Intersections in the triangle-based approach with triangles sharing an edge}
   \label{fig:intersections_triangulation_halfspace}
 \end{figure}

There are other intersections possible besides those in which two triangles share an edge.
Two triangles $\triangle_i, \triangle_j$ that share an edge $e$ intersect if both remaining points
lie on the same side of $e$.
Figure~\ref{fig:intersections_triangulation_halfspace} illustrates this idea
for three triangles. Points $s_i,s_j \in S$ lie on the same side of $e$.
Therefore, the triangles they form with $e$ do intersect. For points on the
opposite site of $e$, such as $s_k$, the induced triangles with $e$ cannot
intersect with those from $s_i,s_j$.

For every possible edge $e$ of the triangulation of the optimal
polygon, at most one triangle may be on each side of the edge. We
refer to $e$ as a hyperplane dividing the two-dimensional space into two halfspaces.
Each halfspace may have one triangle $\triangle$ containing $e$.
This allows us to formulate \emph{halfspace constraints}, as follows.

\begin{equation}\label{eq:halfspace_constraints}
\forall e=\{s_i,s_j\} \text{ with } s_i \neq s_j \land s_i,s_j \in S:
\begin{array}{l}
\sum_{\subscripttriangle \in H^+(e)} x_\subscripttriangle \leq 1\\
\sum_{\subscripttriangle \in H^-(e)} x_\subscripttriangle \leq 1
\end{array}
\end{equation}

$H^+$ and $H^-$ are the two halfspaces induced by $e$ (see Figure~\ref{fig:intersections_triangulation_halfspace}).
Because chords of the convex hull (CH-chords) cannot lie on the boundary of the polygonization,
we can use the \emph{halfspace constraints} to formulate a condition that applies to triangles that
contain such edges. For these triangles, the sum over both halfspaces
has to be the same to ensure that either the triangles are not part of
the solution or if they are, there has to be a triangle on both sides of the
edge.

\begin{equation}
\forall \text{CH-chords } e:
\sum_{\subscripttriangle \in H^+(e)} x_\subscripttriangle = \sum_{\subscripttriangle \in H^-(e)} x_\subscripttriangle
\end{equation}

\subsubsection{Branching on Variables}\label{sec:branch_on_variables}

Another fine-tuning trick is based on a simple concept that makes use of the CPLEX
callback API. As stated before, we may have not added all intersection constraints
from the beginning. This leads to many interim solutions with intersecting
edges.

Imagine a branching where two branches are created; subtree $T_1$ sets
$x_i=1$ and $T_0$ sets $x_i=0$. We are interested in the branch $T_1$. In
that subtree, $x_i$ is set to one. For all child nodes of $T_1$, object $o_i$ is
part of the solution. Therefore, all intersecting object may not be set to
one. In order to prevent the solver from unnecessary branching, we set all the
intersecting entities to zero when branching on variable $x_i$. This leads to
fewer intersection constraints and less branching in later stages.

For the triangle-based approach, we can make use of another
characteristic. If we are certain that a group of triangles will not be part of
a solution in the current branch, these triangles may be the last ones that
are able to connect two unconnected components. In case we already branched
variables of one component to one, we can exclude all variables of the other
component. If we branched variables of both components to one, we can prune the
current node, because there will be no future solution which connects both
components.

\subsubsection{Subtour Constraints} \label{sec:subtour_constraints}
Both integer program formulations make use of the concept of \emph{subtour
constraints}. As described, these are only added incrementally during the
optimization process.

\paragraph{Callback Graphs}
The computation of subtours is problem specific in certain areas. However, if
we abstract both problem's interim solutions on undirected callback graphs $G$,
the algorithmic approaches have multiple attributes in common. For the edge-based
approach we create a vertex for each point $s_i \in S$ and
connect those vertices $v_i,v_j$, where $(i,j)$ or $(j,i)$ are part of the
solution. In the triangle-base approach we build the dual graph, i.e., we have
a vertex for each triangle of the solution, and an edge between adjacent triangles.

\paragraph{Connected Components}
Finding violated subtour constraints in a given interim solution can simply be based
on searching for connected components in the callback graph.
For computing multiple connected component at once, we use a DFS-based approach that
starts at a vertex $v_i$ and finds its connected component by iterating through
the edges of the callback graph. If there is any non-visited vertex $v_j$, it
repeats the last step until all vertices were visited. This method operates in
time $O(n + \vert E \vert)$.

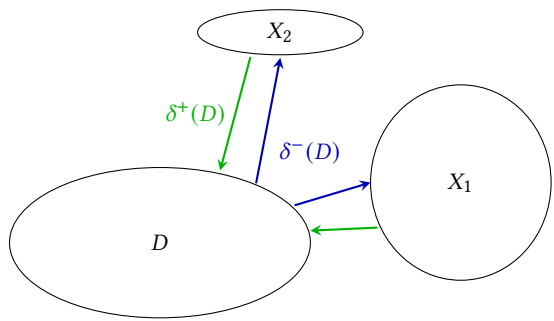
\begin{figure}[h]
	\centering
	\begin{tikzpicture}[line cap=round,line join=round,>=stealth,x=.4cm,y=.4cm]
\definecolor{lightblue}{RGB}{50,89,200}
\definecolor{lightgreen}{RGB}{45,200,76}

 \draw (0,0) ellipse (2cm and 1cm) node[] {$D$};
 \draw (10,2) ellipse (1.2cm and 1.3cm) node[] {$X_1$};
\draw (4,7) ellipse (1.1cm and 0.3cm) node[] {$X_2$};

\draw[-stealth, thick, black!30!blue] (4.5, 1.25) -- (7,2);
\draw[-stealth, thick, black!30!blue] (3.2, 2) -- (4,6.15);
\draw[black!30!blue] (5,3) node[] {$\delta^-(D)$};

\draw[stealth-, thick, black!30!green] (5, 0.4) -- (7.2,0.5);
\draw[stealth-, thick, black!30!green] (2, 2.4) -- node[left] {$\delta^+(D)$} (3,6.15);
\end{tikzpicture}
	\caption{Three connected components of an interim solution}
	\label{fig:connected_components:reference_point_approach}
\end{figure}

\paragraph{Edge-Based Approach}
For a given interim solution we compute the connected components;
for each such component $D$, we add one constraint over the sum of outgoing edges of $D$ and one constraint
over the incoming edges of $D$.
\begin{equation*}\sum_{e \in \delta^-(D)} z_e \geq 1 \qquad \sum_{e \in \delta^+(D)} z_e \geq 1\end{equation*}
 The constraints can be generated in $O(\vert E \vert)$, because we need to
iterate over all edges to find out which one are leaving or entering
the component. Figure~\ref{fig:connected_components:reference_point_approach}
illustrates an example of three connected components in a solution. 
Observe that one of $\delta^+$ and one of $\delta^-$ edges has to be part of a
correct solution in order to connect $D$ to the rest of the point set.

\paragraph{Triangle-Based Approach}
For a given interim solution we compute the connected components. In contrast
to the edge-based approach, we cannot find two sets $\delta^+,\delta^-$,
because the vertices of the callback graph do not have to be part of the
optimal solution. This makes the problem of preventing subtours more complex,
because each subtour constraint has to include an information about which
vertices have been chosen so far. The main idea is to force a given component
to have at least one neighbor included in the optimal solution. Iterating these
constraints over the interim solutions will force two unconnected components to
get connected.


Let $D$ be a connected set of triangles and let $\delta(D)$ be the set of all
triangles having one edge on the outer boundary of $D$ and one point in
$T\setminus D$.
\begin{equation}
   \sum_{\subscripttriangle \in D} x_\subscripttriangle \leq \vert D \vert - 1 + \sum_{\subscripttriangle \in \delta(D)} x_\subscripttriangle
   \label{eq:triangulation_approach:simple_subtour_constraint}
 \end{equation}
 Observe that the constraint forces a solution containing $D$ to attach at least one triangle to $D$.
 Note that the constraint is not as strong as the intersection constraint of the edge-based approach, because it depends on the configuration of $D$.
 The constraint is also satisfied if one triangle of $D$ gets exchanged by another.

\paragraph{Finding Minimum Cuts}
If we consider fractional solutions of the Linear Programming relaxations of either problem,
finding violated subtour constraints requires searching for \emph{minimum cuts}.
The simplest approach is to use one of the well-known max.flow algorithms.
Ford and Fulkerson~\cite{ford1956maximal} gave an elegant proof that a maximum flow $f$ is also a minimal cut in a flow
network. For undirected graphs, Stoer and Wagner~\cite{stoer1997simple}
provided an algorithm that finds a minimal cut in $O(\vert V \vert \cdot \vert E \vert
+ \vert V \vert^2 \log \vert V \vert)$. To find more than one subtour,
we can also search for min-cuts separating two given vertices $v_i,v_j$. A
minimum cut of $G$ will be one of the cuts separating two vertices $v_i \in
V_1$ and $v_j \in V_2$. Gomory and Hu~\cite{gomory1961multi} introduced the
concept of edge-weighted \emph{Gomory-Hu-Trees} $T(G)$. For every pair of
vertices $v_i,v_j \in V(G)$ the minimum cut separating $v_i$ and $v_j$ in $G$
is the minimum weighted edge of the path from $v_i$ to $v_j$ in $T(G)$. Using
this technique we are able to add more than one subtour constraint at once.

\subsection{Subtour Angle Constraints}\label{sec:subtour_angle_constriant}
For the next idea in the triangle-based approach we consider possible subtours.

\begin{figure}[b]
  \centering
  \includegraphics[scale=0.7]{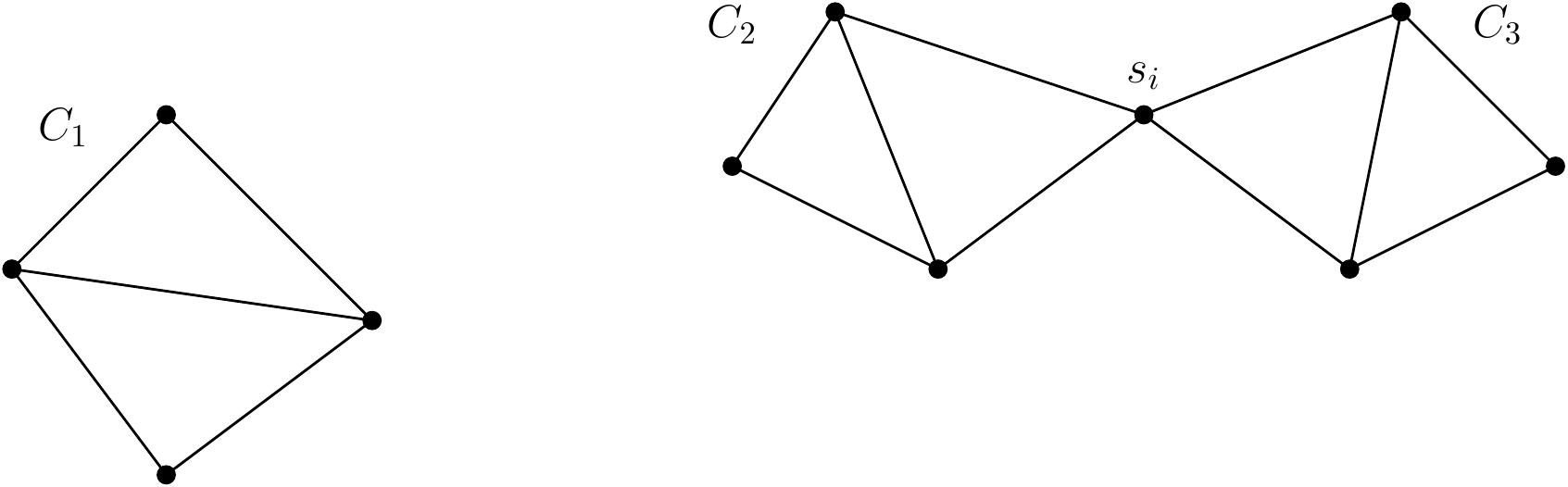}
  \caption{Subtour types in the triangle-based approach: $C_1$ does not share a point with $C_2$ or $C_3$; $C_2$ and $C_3$ share the point $s_i$.}
  \label{fig:subtour_types_triangulation_approach}
\end{figure}

Figure~\ref{fig:subtour_types_triangulation_approach} illustrates two different
types of subtours that may occur in interim solutions of the triangle-based
approach. The first type are connected components in the sense that two
components $C_i, C_j$ do not share a single point. In the illustration, $C_1$
and $C_2$ as well as $C_1$ and $C_3$ are components of such type. The second
type are components that share one or more points. In
Figure~\ref{fig:subtour_types_triangulation_approach}, this type is represented
by the components $C_2$ and $C_3$ that share a point $s_i \in S$.

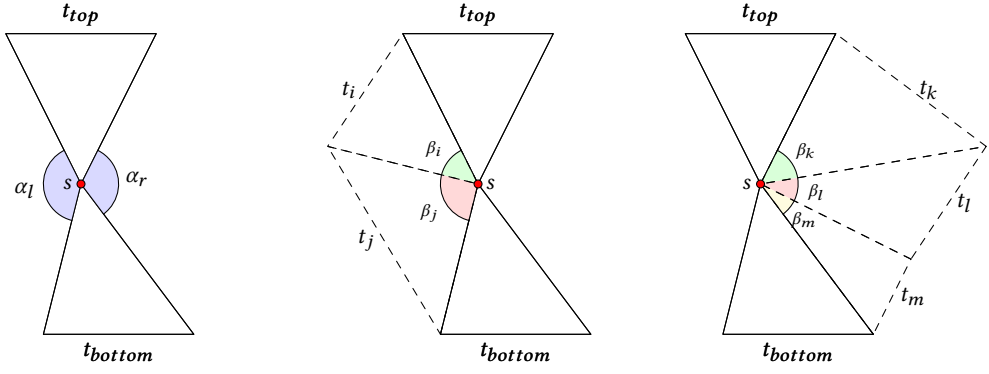
\begin{figure}[h]
  \centering
  \begin{subfigure}[b]{.3\textwidth}
\centering
\begin{tikzpicture}[line cap=round,line join=round,>=stealth,x=1cm,y=1cm]

\draw [] (1,2) coordinate (ar1) --  (0,0) coordinate (zero) -- (-1,2) coordinate (al1) -- node[above] {$t_{top}$} cycle ;
\draw [] (1.5,-2) coordinate (ar2) --  (zero) -- (-.5, -2) coordinate (al2) -- node[below] {$t_{bottom}$} cycle;

\pic [draw, -, fill=blue!15, "$\alpha_r$", angle eccentricity=1.5] {angle = ar2--zero--ar1};
\pic [draw, -, fill=blue!15, "$\alpha_l$", angle eccentricity=1.5] {angle = al1--zero--al2};

\draw [] (1,2) coordinate (ar1) --  (0,0) coordinate (zero) -- (-1,2) coordinate (al1) -- node[above] {$t_{top}$} cycle ;
\draw [] (1.5,-2) coordinate (ar2) --  (zero) -- (-.5, -2) coordinate (al2) -- node[below] {$t_{bottom}$} cycle;

\draw [fill=red] (0,0) circle (1.5pt) node[left] {$s$};

\end{tikzpicture}
\caption{Point connecting two components}
\label{fig:fan_expressions:two_components}
\end{subfigure}\hfil
\begin{subfigure}[b]{.3\textwidth}
\centering
\begin{tikzpicture}[line cap=round,line join=round,>=stealth,x=1cm,y=1cm]

\draw [] (1,2) coordinate (ar1) --  (0,0) coordinate (zero) -- (-1,2) coordinate (al1) -- node[above] {$t_{top}$} cycle ;
\draw [] (1.5,-2) coordinate (ar2) --  (zero) -- (-.5, -2) coordinate (al2) -- node[below] {$t_{bottom}$} cycle;

\coordinate (t1) at (-2, 0.5);

\begin{scriptsize}
\pic [draw, -, fill=green!15, "$\beta_i$", angle eccentricity=1.5] {angle = al1--zero--t1};
\pic [draw, -, fill=red!15, "$\beta_j$", angle eccentricity=1.5] {angle = t1--zero--al2};
\end{scriptsize}

\draw [] (1,2) coordinate (ar1) --  (0,0) coordinate (zero) -- (-1,2) coordinate (al1) -- node[above] {$t_{top}$} cycle ;
\draw [] (1.5,-2) coordinate (ar2) --  (zero) -- (-.5, -2) coordinate (al2) -- node[below] {$t_{bottom}$} cycle;

\draw [dashed] (al1) --  (zero) -- (t1) -- node[left] {$t_i$} cycle;
\draw [dashed] (al2) --  (zero) -- (t1) -- node[left] {$t_j$} cycle;

\draw [fill=red] (0,0) circle (1.5pt) node[right] {$s$};

\end{tikzpicture}
\caption{Possible solution closing the left side}
\label{fig:fan_expressions:left_side}
\end{subfigure}\hfil
\begin{subfigure}[b]{.33\textwidth}
\centering
\begin{tikzpicture}[line cap=round,line join=round,>=stealth,x=1cm,y=1cm]

\draw [] (1,2) coordinate (ar1) --  (0,0) coordinate (zero) -- (-1,2) coordinate (al1) -- node[above] {$t_{top}$} cycle ;
\draw [] (1.5,-2) coordinate (ar2) --  (zero) -- (-.5, -2) coordinate (al2) -- node[below] {$t_{bottom}$} cycle;

\coordinate (t1) at (3, 0.5);
\coordinate (t2) at (2,-1);
\begin{scriptsize}
\pic [draw, -, fill=green!15, "$\beta_k$", angle eccentricity=1.5] {angle = t1--zero--ar1};
\pic [draw, -, fill=red!15, "$\beta_l$", angle eccentricity=1.5] {angle = t2--zero--t1};
\pic [draw, -, fill=yellow!15, "$\beta_m$", angle eccentricity=1.5] {angle = ar2--zero--t2};

\end{scriptsize}

\draw [] (1,2) coordinate (ar1) --  (0,0) coordinate (zero) -- (-1,2) coordinate (al1) -- node[above] {$t_{top}$} cycle ;
\draw [] (1.5,-2) coordinate (ar2) --  (zero) -- (-.5, -2) coordinate (al2) -- node[below] {$t_{bottom}$} cycle;

\draw [dashed] (ar1) --  (zero) -- (t1) -- node[right] {$t_k$} cycle;
\draw [dashed] (t2) --  (zero) -- (t1) -- node[right] {$t_l$} cycle;
\draw [dashed] (ar2) --  (zero) -- (t2) -- node[right] {$t_m$} cycle;

\draw [fill=red] (0,0) circle (1.5pt) node[left] {$s$};

\end{tikzpicture}
\caption{Possible solution closing the right side}
\label{fig:fan_expressions:right_side}
\end{subfigure}
  \caption{Visualization of subtour angle constraints}
\end{figure}

The construction of the callback graph will detect both component
types, as subtours and equations~\eqref{eq:triangulation_approach:simple_subtour_constraint}
would ensure later solutions that consist of one connected component. We consider two triangles
that share a point $s_i$, but are not connected in the callback graph. If both
triangles are part of the optimal solution, we can be sure that both
triangles are connected with other triangles containing $s_i$. Let $t_{top},
t_{bottom}$ be two triangles of different components sharing one point $s\in
S$. Let $\alpha_l,\alpha_r$ be the angles between the inner edges of
$t_{top},t_{bottom}$ and $s$. Figure~\ref{fig:fan_expressions:two_components}
illustrates both triangles and their respective angles. From now on we assume
that both triangles are part of an optimal solution. Because an optimal
solution has to be a valid polygon, both $t_{top}$ and $t_{bottom}$ have to be
in the same component. The resulting polygon may not have both components only
connected with triangles not including $s$. This leads to the fact that both
components have to be connected via triangles in $\alpha_l$ or $\alpha_r$.
Figures~\ref{fig:fan_expressions:left_side} and
\ref{fig:fan_expressions:right_side} show two possible cases of the optimal
polygon, one closes $\alpha_l$, while the other one closes $\alpha_r$. The sum of
inner angles at $s$ of these triangles has to be equal to $\alpha_l$ and
$\alpha_r$ respectively.
$$\beta_i + \beta_j = \alpha_l \qquad \beta_k + \beta_l + \beta_m = \alpha_r$$
Note that no solution can have both angles filled completely, because $s$ would
no longer be on the boundary of $P$. This leads to
so-called \emph{subtour angle constraints}. The main idea is that if both
triangles $t_{top},t_{bottom}$ are part of the solution, other triangles at $s$
need to close at least an angle of $\min \{\alpha_l, \alpha_r\}$.

\begin{equation}\label{eq:subtour_angle_constriant}
  \min \{\alpha_l,\alpha_r\} \cdot (z_{top} + z_{bottom} - 1) \leq \sum_{\subscripttriangle \in \delta(s)} \beta_{\subscripttriangle}^s \cdot x_\subscripttriangle
\end{equation}

By $\beta_\subscripttriangle^s$ we denote the inner angle of the triangle
$\triangle$ at point $s$. Figures~\ref{fig:fan_expressions:left_side} and
\ref{fig:fan_expressions:right_side} illustrate how the angles
$\beta_\subscripttriangle^s$ add up to $\alpha_{l/r}$. At every integer
solution we obtain during the solving process, we have reason to believe that
triangles of the integer solution have a high probability to be part of the
optimal solution. Because of that, we add \emph{subtour angle constraints} at
every integer solution. In addition to the previous ideas, we generalize the
idea of two triangles at one point to so-called \emph{triangle fans}.

\begin{definition}
  In an interim solution a \emph{triangle fan} is a set $F_s$ of connected triangles $\triangle$, which share a point $s \in S$. This means that there is a path $P$ for every $\triangle_i, \triangle_j \in F_s$ in the callback graph.
\end{definition}

Consider an interim solution obtained during the solving process. 
After constructing the callback graph, we iterate over all
triangles of the solution and add each triangle to three triangle fans (one
for each point of $\triangle$). The triangle $\triangle$ will be added to an
existing fan $F_s$ if it is in the same component, because this indicates a
path between $\triangle$ and the other $\triangle_i \in F_s$. If no such fan is
found, a new fan will be created for this component. In the end
we return all triangle fans for each point $s\in S$. 
Because a solution contains $n-2$ triangles, we will add at most $n-2$ triangles to their fan. 
As the triangle has three points, each triangle can be part of three fans. 
This leads to a time complexity of $O((n-2) \cdot 3 \cdot (n-2))=O(n^2)$ where $n$ is the number of points in $S$.

%
%

After computing all triangle fans we choose two triangles
$t_{top},t_{bottom}$ with a common point $s$ from two different fans in order to
formulate the constraint. 
These triangles should not be chosen at random, because
the constraint gets much stronger if the following two conditions apply.

\begin{itemize}
  \item The triangles should have a small area.
  \item The angles $\alpha_l,\alpha_r$ are similar, i.e., they minimize $\vert \alpha_l-\alpha_r \vert$.
\end{itemize}

Consider two maximal fans $F_1, F_2$ having a point $s$ in common.
We choose triangles $t_{top}\in F_1,
t_{bottom}\in F_2$, such that $\vert \alpha_l-\alpha_r \vert$ is minimized. 
If there are multiple candidates, 
we choose those that minimize the area.

\subsubsection{Point-based Subtour Constraints}
\begin{figure}
  \centering
  \includegraphics[page=1, scale=.6]{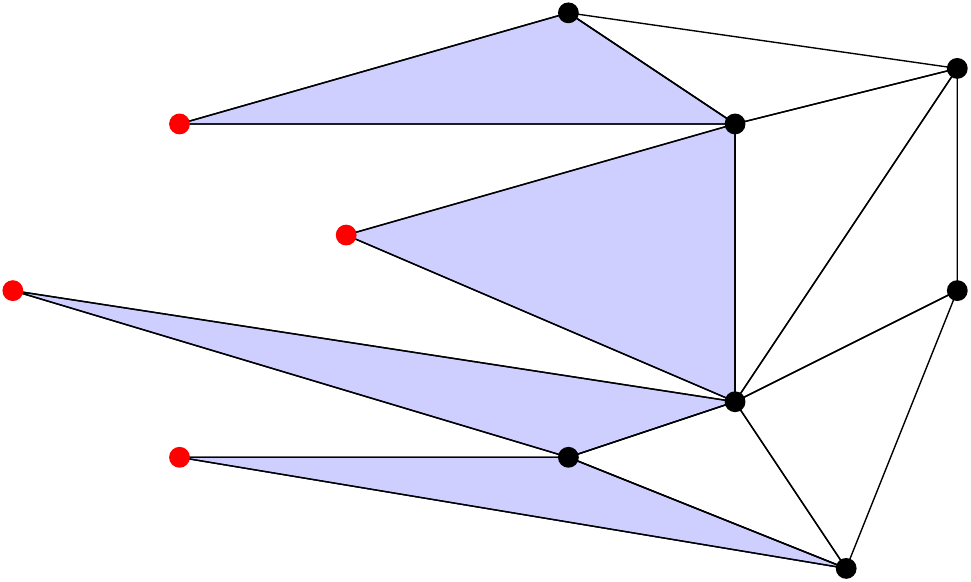}
  \hfil
  \includegraphics[page=2, scale=.6]{figures/point_based_subtour_constraint.pdf}
  \caption{Illustration of point-based subtour constraints. 
  	Red points correspond to the set $X$.
  	Left: A valid solution when no outgoing triangle (blue triangles) has two points from $X$. 
  	Then there must be $\vert X\vert$ triangles connecting $X$ to $S\setminus X$.
	Right: If at least one triangle has two points from $X$, then there must exist at least one more outgoing triangle.}
  \label{fig:pointbased_subtour}
\end{figure}

Consider a point set $X \subseteq S$ of an interim solution with $3 \leq \vert X
\vert \leq n-2$. We denote $\triangleone \in \delta(X)$ and $\triangletwo \in
\delta(X)$ to be triangles with exactly one corner or two corners inside $X$
respectively. We know that the point set $X$ has to be connected to at least
two points outside of $X$. To connect both points to $X$, at least two triangles
are needed that have at least one point outside of $X$ (see Fig.~\ref{fig:pointbased_subtour} right). This leads to the
first point-based subtour constraint
\begin{equation}\label{eq:point_based_subtour_one}
\sum_{\subscripttriangleone \in \delta(X)} x_{\subscripttriangleone} + \sum_{\subscripttriangletwo \in \delta(X)} x_{\subscripttriangletwo} \geq 2
\end{equation}

Now suppose there is no triangle connecting two points from $X$.
This implies that each point $s\in X$ needs a triangle connecting $s$ with two points from $S\setminus X$.
Therefore, if $\sum_{\subscripttriangletwo \in \delta(X)} = 0$, then $\sum_{\subscripttriangleone \in \delta(X)} x_{\subscripttriangleone} = \vert X\vert$ (see Fig.~\ref{fig:pointbased_subtour} left).
If there is at least one triangle with two points in $X$, then a possible solution can exist with only one additional triangle.
Thus, if $\sum_{\subscripttriangletwo \in \delta(X)} x_{\subscripttriangletwo} \geq 1$, then $\sum_{\subscripttriangleone \in \delta(X)} x_{\subscripttriangleone} +\sum_{\subscripttriangletwo \in \delta(X)} x_{\subscripttriangletwo} \geq 2$ (see Fig.~\ref{fig:pointbased_subtour} right).
Combining both cases yields the following constraint for a point
set $X$.

\begin{equation}\label{eq:point_based_subtour_two}
\sum_{\subscripttriangleone \in \delta(X)} x_{\subscripttriangleone}  \geq \vert X \vert - (\vert X \vert - 1) \cdot \sum_{\subscripttriangletwo \in \delta(X)} x_{\subscripttriangletwo}
\end{equation}

Separation over these constraints can be achieved analogous to regular subtour constraints
for the classic TSP, with triangles in our problem corresponding to vertices in the TSP,
and connected components corresponding to connected sets of triangles. This allows
polynomial-time separation, but requires iterating over the $O(n^3)$ triangles.

	\section{Experiments}
\label{sec:experiments}

Based on the described approaches, we ran experiments on some machines with slightly different specifications and parameters.
We used CPLEX 12.9 with a time limit of 1800 seconds on an AMD Ryzen 7 5800X CPU \@ 4.2GHz with eight cores and 16 threads utilizing an L3 Cache with a size of 32MB.
The solver was able to use a maximum amount of 128GB RAM.
Our solver uses the default CPLEX parameters except \textsc{CPXPARAM\_Parallel}, which was set to \textsc{CPX\_PARALLEL\_OPPORTUNISTIC}.

We considered all instances from the CG:SHOP Challenge with up to 50 points;
see Section~\ref{subsec:cgshop} for a detailed description.  Because the
original CG:SHOP benchmark set mostly aims at heuristic and experimental
methods developed in the competition, it reaches all the way up to 1,000,000
points, but is relatively sparse within the range of exact methods.  We
accounted for this sparsity by generating additional instances of similar type;
because of fast run times, we used 20 instances and 5 iterations each for
instance sizes 12-20; for instance sizes 21-23, we considered 20 instances and
1 iteration each; for larger sizes, we limited the number to 10 instances each.

\subsection{Solver Types}

In this section we will introduce different versions for each solver that implement features that we mentioned in Section~\ref{sec:tools}. For both the triangulation-based and edge-based approach, we pass a start solution that was generated by a \textsc{Greedy \textsc{Min-Area}} heuristic that is inspired by the work of Taranilla et al.~\cite{taranilla2011approaching},
Fekete's \textsc{Max-Area} approximation~\cite{f-gtsp-92} or solutions from the CG:SHOP competition. \textsc{Greedy \textsc{Min-Area}} starts with a polygon $P=conv(S)$ and carves out the largest triangles by replacing an edge $(p_i,p_j)$ of $P$ by two edges $(p_i, q)(q, p_j)$ to an inner point $q$.

\subsubsection{Edge-Based Solvers}
\textsc{EdgeV1} is a basic integer program of the edge-based approach.
It adds all intersection constraints and slab constraints before starting the solving process and
adds subtour constraints in every integer solution. This integer program is an
improvement to the edge-based \textsc{MinArea} integer program presented by
Papenberg et al. \cite{Papenberg2014, fekete2015area}.
In the former approach
cycle based subtour constraints were added after an optimal solution has been
found. This resulted in poor computing times even for small point sets.
We also utilize properties of the convex hull to exclude certain variables, i.e., edges that connect two
non-adjacent points on the convex hull, from the computation.
\textsc{EdgeV1} makes use of this concept by setting these variables to
zero.
\textsc{EdgeV2} extends the previous version by adding intersection
constraints at interim solutions.
\textsc{EdgeV3} includes a branching extension where branching on a variable $z_e$ results in intersecting edges getting branched to zero.
In \textsc{EdgeV4} we additionally search for subtours in fractional interim
solutions and add slab constraints during the solving process. The upcoming sections will show that the edge-based approach is better suited for \textsc{Max-Area} instances.

\subsubsection{Triangle-Based Solvers}

\textsc{TriangulationV1} is the first version of the triangle-based approach.
Compared to the basic triangulation approach of Papenberg \cite{Papenberg2014},
we have fewer variables and different subtour constraints
\eqref{ip:triangulation_approach:subtours}. We added further \emph{halfspace inequalities}
as well as equalities for edges which connect
non-adjacent vertices of the convex hull.
In \textsc{TriangulationV1} we add subtour constraints and intersection
constraints in every integer solution. \textsc{TriangulationV2} extends the
first version with so-called \emph{subtour angle constraints}.
These are added
at every integer solution. We are able to reuse the connected components we need to compute
along the way.
This allows us to add constraints~\eqref{ip:triangulation_approach:subtours} without much additional computation
time. \textsc{TriangulationV3} makes use of additional results on ineffective subtour constraints.
In addition to the constraints of
\textsc{TriangulationV2}, we add point-based subtour constraints
to every intermediate integer solution. The upcoming sections will show that the triangulation-based approach is better suited for \textsc{Min-Area} instances.

\subsection{Analysis}
\label{sec:computation-time}

\begin{figure}[h]
  \includegraphics[width=.9\linewidth]{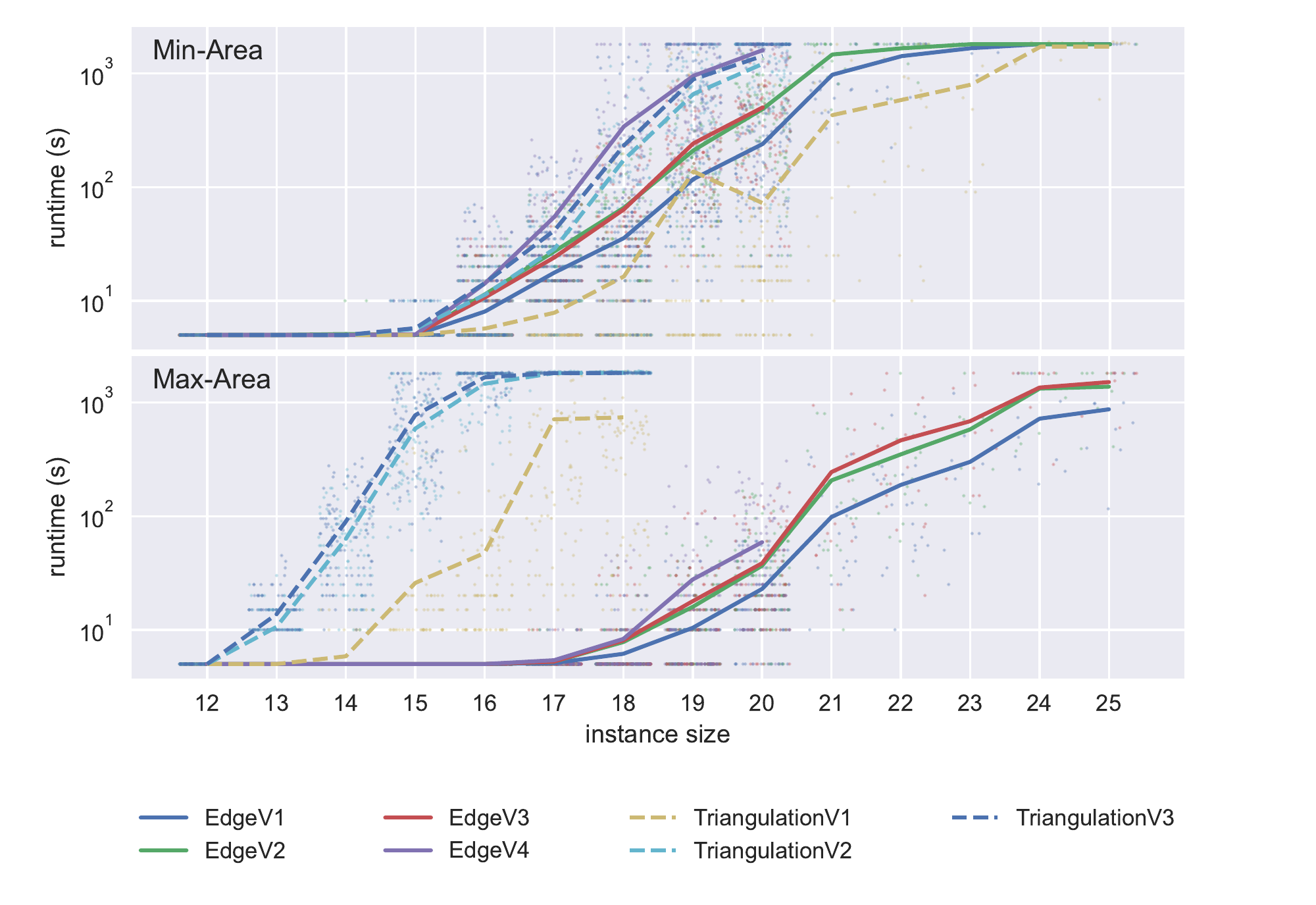}
  \caption{Runtimes for different solver versions on random instances of size $12-25$ and a time limit of $1800$ seconds. The line is the average runtime over five iterations on $20$ instances for sizes $12-20$. For sizes $21-25$ we solved one iteration on $20$ instances for sizes $21-23$ and $10$ instances of size $24-25$.}
  \label{fig:random-detailed-runtime}
\end{figure}

\begin{figure}[h]
	\includegraphics[width=.9\linewidth]{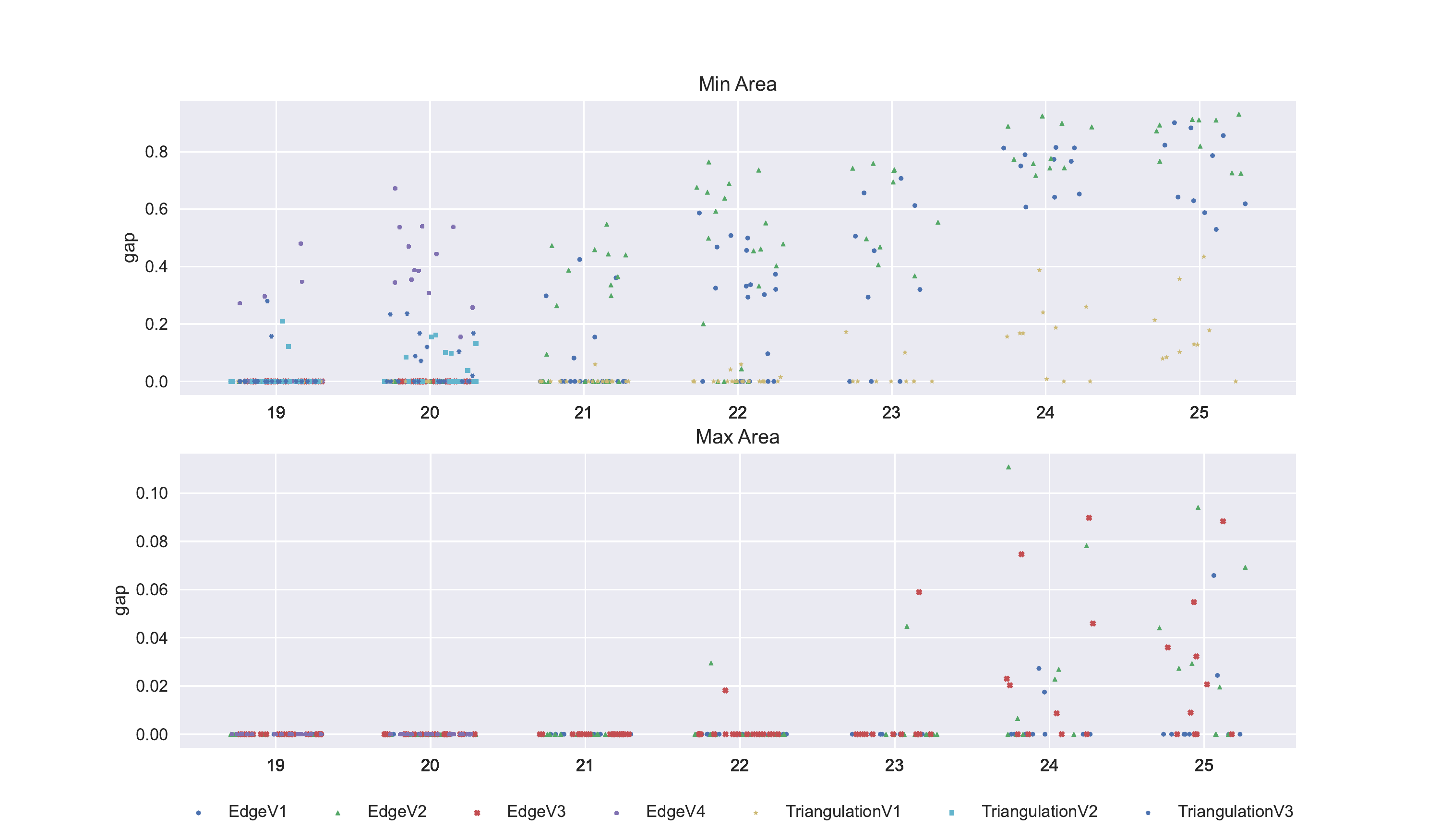}
	\caption{Optimality gap for instances with 19 to 25 points. Shown is the best gap over multiple runs for each instance.}
	\label{fig:random-detailed-gap}
\end{figure}

\begin{figure}[h]
  \includegraphics[width=.9\linewidth]{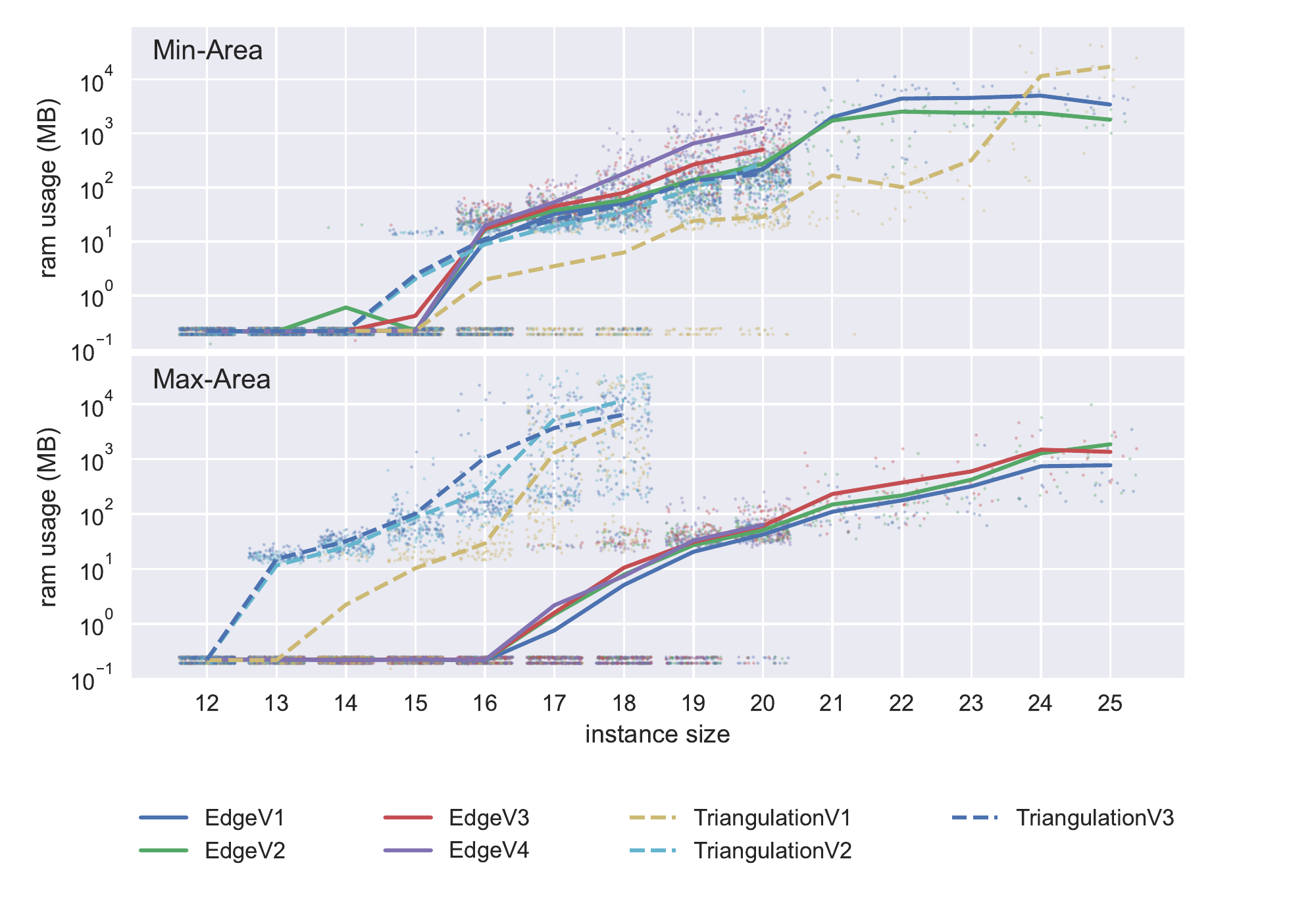}
  \caption{RAM usage for the experiments in Figure~\ref{fig:random-detailed-runtime}.}
  \label{fig:random-detailed-ram-usage}
\end{figure}

\subsubsection{Edge-Based Solvers}

We compared two different approaches. The
first one, adds intersection constraints at every
integer and at every fractional solution. The second one, adds intersection constraints at every integer solution.
Our observations showed that searching for intersections in fractional solutions
increases the computation time. We assume that the intersection constraints
we obtain from fractional solutions are not needed for computing the optimal
solution and that their generation wastes a lot of computation time. We denote
\textsc{EdgeV2} as the version which adds intersection constraints in every
integer solution.

Figure~\ref{fig:random-detailed-runtime} provides a detailed view on the
runtime of the major \textsc{Edge} versions we consider in our experiments for instances of size $12-25$. The scatter plot shows the runtimes for different iterations and instances while the line is the average runtime over five iterations and $20$ instances for each $n \leq 20$. For $n > 20$, we computed one iteration on $20$ instances (for sizes $21-23$) and $10$ instances (for sizes $24-25$). 
Figure~\ref{fig:random-detailed-gap} provides an overview over the optimality gap for instances with at least 19 points. 
Figure~\ref{fig:random-detailed-ram-usage} illustrates the amount of memory that was used during the execution.
As in other problems like TSP, adding constraints in interim solutions instead of adding all constraints in the beginning will only make significant impact when the instances reach a certain size.
Our goal is to add fewer constraints in the beginning while preserving the baseline runtime and the high success rate of \textsc{EdgeV1} for smaller instances.
The best approach can then be used to solve larger instances where the construction of the complete integer program consumes too much space.

In comparison to the first version, \textsc{EdgeV2} adds intersection
constraints at every integer solution. Whenever many intersections constraints were needed to obtain the optimal solution, we observed that the runtime and the depth of the branch and bound tree increased. Figure~\ref{fig:random-detailed-runtime} shows that the runtime for this approach was slightly higher than the \textsc{EdgeV1} baseline.
\textsc{EdgeV3} further adds intersection constraints during branching, which preserved the low runtime on some instances while mitigating the negative effect on instances that needed a lot of intersection constraints.
Apart from intersections one might want to add the slab constraints in interim solutions. 
Slab constraints ensure that the resulting polygon is oriented in the right direction. 
\textsc{EdgeV4} adds slabs constraints during execution and searches for connected components, i.e. min-cuts in every fractional solution. 
We noticed that almost all of the slab constraints are added to the IP at some point during the execution of \textsc{EdgeV4}. 
As a consequence these constraints should be added from the beginning. 
In our implementation of \textsc{EdgeV4}, we computed the connected components of the interim solutions, i.e., minimum cuts with value one. 
We implemented multiple versions for finding minimum cuts of greater size. 
All approaches deteriorated the time needed for the computation. 
Moreover, the approach was unable to solve all instances of larger sizes within the time constraint. 
This shows that the computation of larger cuts is computationally expensive and the obtained inequalities not worth the effort.

Despite the low runtime on smaller instances, the number of unneeded intersection
constraints added by \textsc{EdgeV1} grows fast for larger instances.
Therefore, the other approaches add fewer constraints, which results in better computation times.
Overall the \textsc{EdgeV2} approach which adds intersections in integral interim solutions or \textsc{EdgeV3} which further uses some advanced branching techniques appear to be the best approaches for solving larger instances.

\subsubsection{Triangle-Based Solver}

As shown in Figure~\ref{fig:random-detailed-ram-usage}, the RAM usage of both approaches  depend on the problem variant we are trying to solve.
In Section~\ref{sec:triangle} we proposed the triangle-based IP which uses $O(n^3)$ variables and $O(n^6)$ constraints (excluding subtour constraints).
As a consequence, the formulation, branch and bound tree and temporary variables of the integer program require a lot of space on the executing machine. This is especially relevant for the \textsc{Max-Area} variant, for which many intersection constraints are needed to obtain a feasible solution.

Figure~\ref{fig:random-detailed-runtime} provides a detailed view on the
runtime of the major \textsc{Triangle} versions we considered in our experiments. Due to the triangulation approach performing worse on the \textsc{Max-Area} variant, we excluded instances of size $\geq 19$ in the experiment.
The scatter plot shows the runtimes for different iterations and instances while the line is the average runtime over five iterations and $20$ instances for each $n \leq 20$. For $n > 20$, we computed one iteration on $20$ instances (for sizes $21-23$) and $10$ instances (for sizes $24-25$).

As described above, there are only few
intersections in interim solutions of the triangle-based approach when solving
\textsc{Min-Area} instances, because overlapping areas are counterproductive for obtaining a
minimal solution. Nevertheless, intersection constraints are possible and
needed to be eliminated if no other constraints were found. The results of the
edge-based approach showed that adding intersection clique constraints can be very efficient.
We adapted the idea and searched for intersection cliques in \textsc{TriangulationV1} as well.
In \textsc{TriangulationV2}, we search for subtour angle constraints as well.
Subtour angle constraints can help to decrease the runtime for some instances. In other occasions, the approach that added these constraints performed worse. We assume that better results can be achieved, if one improves the algorithm that finds the possible subtour constraints. We did not further investigate this opportunity as the triangulation approach has large space requirements for larger instances. Figure~\ref{fig:random-detailed-ram-usage} shows that \textsc{TriangulationV1} has a higher RAM usage than the edge-based approaches for $n\geq 24$. When attempting to solve the CG:SHOP instances to optimality in the next section, we reached the limit of $128GB$ for some instances of size $\geq 30$ and all instances of size $\geq 40$. The point based subtour constraints that were added in \textsc{TriangulationV3} did not help to improve the performance of the approach and should not be considered as a valuable extension.

\subsubsection{Convex Hull}

During our experiments, we noticed a strong relation between the number of points on the convex hull in an instance and the runtime that is needed to find an optimal solution.
We therefore generated instances with $n$ points and $3 \leq k \leq n$ points on the convex hull.
This was done by first generating $k$ points in convex position, taking points uniformly at random,
discarding all points that would either lie in the interior or cause some previously selected points to lie in the interior.
We then added $n-k$ points within the convex hull chosen uniformly at random.

\begin{figure}[h]
  \includegraphics[width=.85\linewidth]{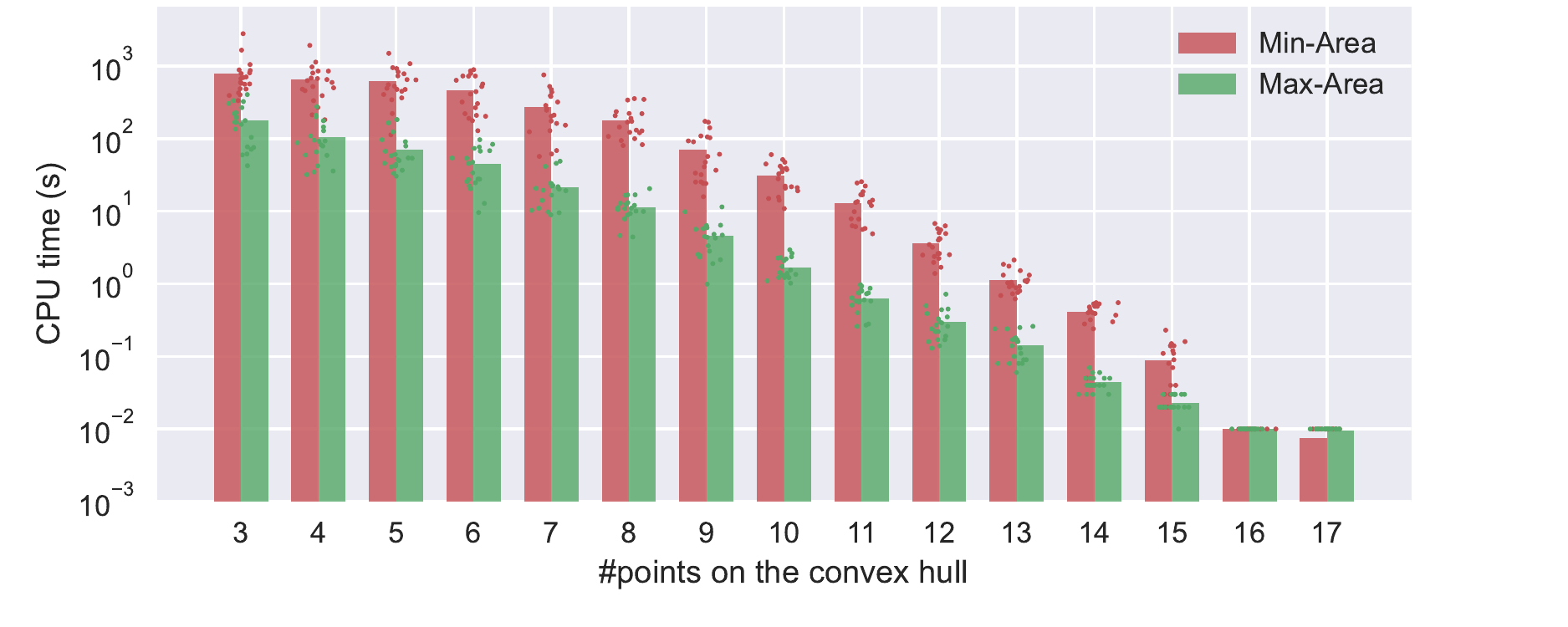}
  \caption{Runtime for the solver \textsc{EdgeV1} with $n=17$ points and $20$ instances with a convex hull size of $k$ for every $k=3,\dots,17$.}
  \label{fig:17-convex-edge}
\end{figure}

\begin{figure}[h]
  \includegraphics[width=.85\linewidth]{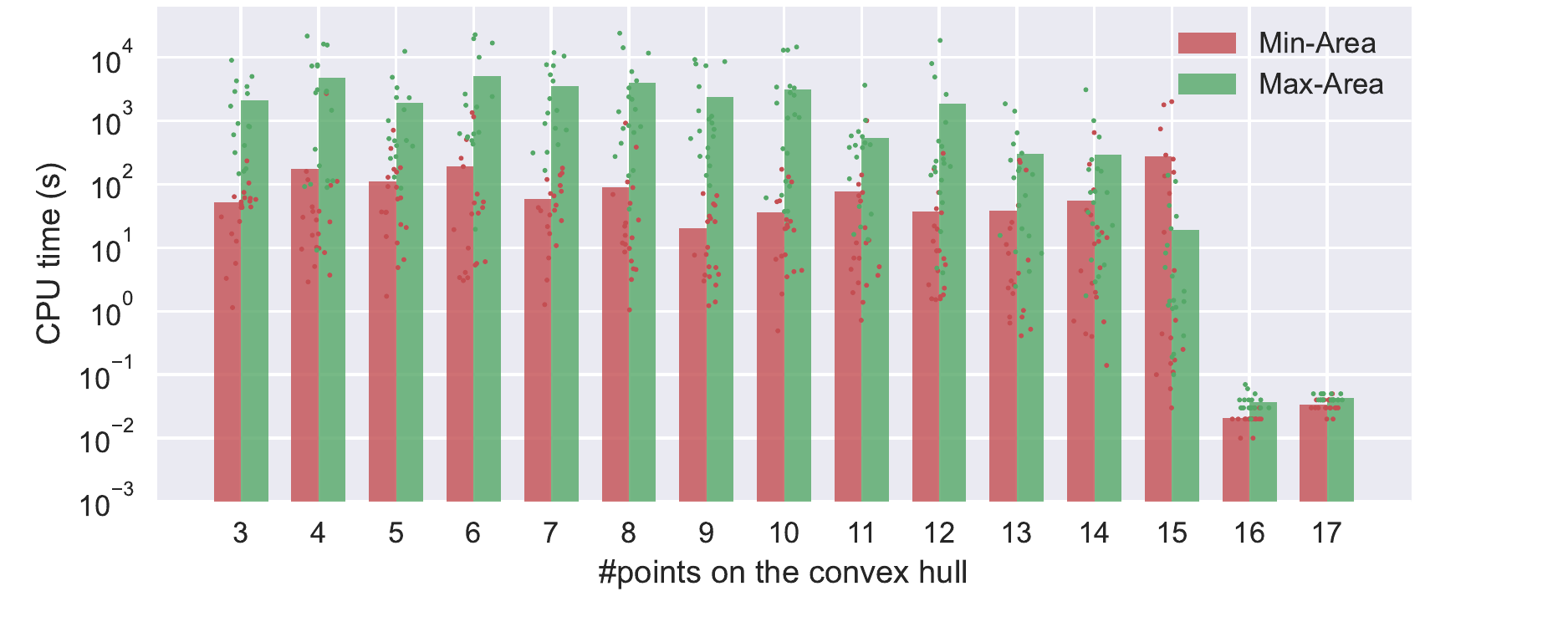}
  \caption{Runtime for the solver \textsc{TriangulationV1} with $n=17$ points and $20$ instances with a convex hull size of $k$ for every $k=3,\dots,17$.}
  \label{fig:17-convex-triangle}
\end{figure}

We performed an experiment for the solver \textsc{EdgeV1} with $n=17$ points and generated $20$ instances for every $k=3,\dots,17$.
Figure~\ref{fig:17-convex-edge} compares the average aggregated CPU times, that is, the sum of time used by all processors during the optimization phase.
We can clearly observe that the CPU time drastically decreases when more points lie on the convex hull of a point set.
We assume that this is due to the fact that fewer edges are possible candidates and the number of possible polygonizations is smaller for these instances.
Moreover, edges that connect two non-adjacent points of $conv(S)$ are set to zero, simplifying many constraints of the IP.
On the other hand, Figure~\ref{fig:17-convex-triangle} shows the average CPU times for the triangulation approach \textsc{TriangulationV1}. Apart from a drastic decrease at $k=16,17$, we could only observe a minor trend towards shorter runtimes for larger $k$.

\subsection{CG:SHOP Results}
\label{subsec:cgshop}

In this section, we discuss the results both IPs obtained on the CG:SHOP competition instances. We start off with small instances that the approaches were able to solve optimally in short time. Table~\ref{tab:cgshop-small-10-15} shows the runtimes for the smallest instances of the competition.\\

\begin{table}[h]
  \footnotesize
  \begin{tabular}{ll|rr|rr}
\toprule
instance & size & {\textsc{Min-Area}} & runtime (s) & {\textsc{Max-Area}} & runtime (s)\\
\midrule
  uniform-0000010-1 &   10 &            58872 &  0.10000 &        148010  &  0.02000\\
  uniform-0000010-2 &   10 &            51568 &  0.03000 &        151540  &    0.03000\\
  london-0000010 &   10 &         19207298 &  0.09000 &      95364394  &  0.04000\\
euro-night-0000010 &   10 &          9954272 &  0.06000 &      28498980  &  0.04000\\
   us-night-0000010 &   10 &          3650560 &  0.10000 &      13059816  &  0.08000\\
      stars-0000010 &   10 &      40264246222 &  0.40000 &  127476507282  &  0.02000\\
uniform-0000015-1 &   15 &           102716 & 14.03000 &        391474  &  0.58000\\
  uniform-0000015-2 &   15 &           113436 &  3.77000 &        374516  &  0.35000\\
     london-0000015 &   15 &         17809096 &  5.07000 &     127729316  &  0.38000\\
   euro-night-0000015 &   15 &          7107334 & 19.49000 &      37961104  &  2.03000\\
   us-night-0000015 &   15 &         16642190 & 29.90000 &      60295830  &  4.49000\\
        stars-0000015 &   15 &      29776831822 & 13.28000 &  142470465062  &  0.92000\\
\bottomrule
\end{tabular}

  \caption{CG:SHOP results for \textsc{Min-Area} and \textsc{Max-Area} for instances of size $<20$.}
  \label{tab:cgshop-small-10-15}
\end{table}

As point sizes $\leq 15$ have been observed to be solvable by both approaches in a very short time period (see Section~\ref{sec:computation-time}), we only show results from the \textsc{Edge} approach. In comparison with the \textsc{Min-Area} runtimes on uniformly distributed random instances in Figure~\ref{fig:random-detailed-runtime},
 the runtimes on the competition instances appear to be similar. Note that the runtimes in the table are the best runtimes, we observed on these instances instead of the average runtime.\\

As explained earlier,
 the space requirements for the triangulation approach prevents experiments with larger instance sizes. Results from Section~\ref{sec:computation-time} imply that the edge-based approach is better suited for the \textsc{Max-Area} variant. In this paragraph, we investigate the results that were obtained on instances of size $20-50$ from the competition. For larger instances, we raised the maximum runtime to $43,800$ seconds because it is very unlikely to find an optimal solution especially for the larger instances.
As other competitors submitted their best solutions on the same instances, we were able to provide the IP with these solutions, i.e. the best solution that was found during the competition. As most solutions are most likely the optimal solution of the corresponding instance, the main objective was to find good bounds within the time constraint. Keep in mind that we only present the best results that we obtained after numerous attempts on solving the instances.

  {
    \footnotesize
    \begin{longtable}{lcrrrrrrr}
\toprule
           instance & solver & size & competition & IP &  best bound & gap & gap (convex hull) & runtime (s) \\
\midrule
\textbf{euro-night-0000020} &      E &   20 &     43453166 &     43453166 &           43453166 &       0 &         0.09910 &      64.24 \\
    \textbf{london-0000020} &      E &   20 &    200514948 &    200514948 &          200514948 &       0 &         0.11300 &     124.84 \\
     \textbf{stars-0000020} &      E &   20 & 216742984910 & 216742984910 &       216742984910 &       0 &         0.05630 &      16.91 \\
 \textbf{uniform-0000020-1} &      E &   20 &       761968 &       761968 &             761968 &       0 &         0.15700 &     100.63 \\
 \textbf{uniform-0000020-2} &      E &   20 &       804730 &       804730 &             804730 &       0 &         0.08490 &      16.85 \\
  \textbf{us-night-0000020} &      E &   20 &     54561746 &     54561746 &           54561746 &       0 &         0.06340 &      13.04 \\
\textbf{euro-night-0000025} &      E &   25 &     50782240 &     50782240 &           50782240 &       0 &         0.10170 &   15925.92 \\
    \textbf{london-0000025} &      E &   25 &    231544442 &    231544442 &          231544442 &       0 &         0.06950 &    2321.83 \\
     \textbf{stars-0000025} &      E &   25 & 237309787430 & 237309787430 &       237309787430 &       0 &         0.07350 &    6398.03 \\
 \textbf{uniform-0000025-1} &      E &   25 &      1320082 &      1320082 &            1320082 &       0 &         0.19820 &   26839.26 \\
 \textbf{uniform-0000025-2} &      E &   25 &      1379588 &      1379588 &            1379588 &       0 &         0.12530 &    1354.81 \\
  \textbf{us-night-0000025} &      E &   25 &     60585868 &     60585868 &           60585868 &       0 &         0.08360 &  1380.2074 \\
         euro-night-0000030 &      E &   30 &     52830622 &     52830622 &           56079768 & 0.06150 &         0.08350 &          - \\
             london-0000030 &      E &   30 &    151703726 &    151703726 &    161269214.27440 & 0.06310 &         0.09240 &          - \\
              stars-0000030 &      E &   30 & 185057287956 & 185057287956 &       197980842032 & 0.06980 &         0.08390 &          - \\
          uniform-0000030-1 &      E &   30 &      1956068 &      1956068 &      2067442.99220 & 0.05690 &         0.08680 &          - \\
          uniform-0000030-2 &      E &   30 &      2309760 &      2309760 &      2402044.10610 & 0.04000 &         0.07960 &          - \\
           us-night-0000030 &      E &   30 &     66217320 &     66217320 &     70264245.99140 & 0.06110 &         0.07540 &          - \\
         euro-night-0000035 &      E &   35 &     45819094 &     45819094 &     48532070.61540 & 0.05920 &         0.07490 &          - \\
    \textbf{london-0000035} &      E &   35 &    215109002 &    215109002 &          215109002 &       0 &         0.08270 &   100.4485 \\
              stars-0000035 &      E &   35 & 222429730998 & 222429730998 &       239177631315 & 0.07530 &         0.09420 &          - \\
          uniform-0000035-1 &      E &   35 &      3234656 &      3234656 &      3467516.96170 & 0.07200 &         0.08160 &          - \\
          uniform-0000035-2 &      E &   35 &      3255396 &      3255396 &      3494212.54550 & 0.07340 &         0.08450 &          - \\
           us-night-0000035 &      E &   35 &     77135624 &     77135624 &           82463136 & 0.06910 &         0.08810 &          - \\
         euro-night-0000040 &      E &   40 &     47910104 &     47910104 &     50634608.01820 & 0.05690 &         0.05970 &          - \\
    \textbf{london-0000040} &      E &   40 &    209674650 &    209674650 &          209674650 &       0 &         0.13340 &    3948.63 \\
              stars-0000040 &      E &   40 & 190177150422 & 190177150422 & 207054663704.00031 & 0.08870 &         0.11460 &          - \\
          uniform-0000040-1 &      E &   40 &      4431360 &      4431360 &            4822510 & 0.08830 &         0.09010 &          - \\
          uniform-0000040-2 &      E &   40 &      4170194 &      4170194 &            4506200 & 0.08060 &         0.09100 &          - \\
           us-night-0000040 &      E &   40 &     68940956 &     68940956 &     72625076.83330 & 0.05340 &         0.05800 &          - \\
\textbf{euro-night-0000045} &      E &   45 &     48214668 &     48214668 &           48214668 &       0 &         0.07190 &    3267.89 \\
             london-0000045 &      E &   45 &    271205760 &    271205760 &    292963145.99330 & 0.08020 &         0.09140 &          - \\
              stars-0000045 &      E &   45 & 245048373286 & 245048373286 & 266232485329.99460 & 0.08640 &         0.08640 &          - \\
          uniform-0000045-1 &      E &   45 &      4759374 &      4759374 &            5231150 & 0.09910 &         0.09910 &          - \\
          uniform-0000045-2 &      E &   45 &      5158094 &      5158094 &            5682168 & 0.10160 &         0.10160 &          - \\
  \textbf{us-night-0000045} &      E &   45 &     77941112 &     77941112 &           77941112 &       0 &         0.02900 &   292.6256 \\
         euro-night-0000050 &      E &   50 &     60399328 &     60399328 &           65414966 & 0.08300 &         0.24920 &          - \\
             london-0000050 &      E &   50 &    231089684 &    231089684 &          250510480 & 0.08400 &         0.08640 &          - \\
              stars-0000050 &      E &   50 & 247712484090 & 247712484090 & 264902562115.99869 & 0.06940 &         0.06940 &          - \\
          uniform-0000050-1 &      E &   50 &      6385168 &      6385168 &      6899665.99120 & 0.08060 &         0.08060 &          - \\
          uniform-0000050-2 &      E &   50 &      7151224 &      7151224 &      7848421.98230 & 0.09750 &         0.09750 &          - \\
           us-night-0000050 &      E &   50 &     79952918 &     79952918 &           80759104 & 0.01010 &         0.04040 &          - \\
\bottomrule
\caption{CG:SHOP results for \textsc{Max-Area} (edge-based) for instances of size $20-50$}
\label{tab:cgshop-gaps}
\end{longtable}

  }

Table~\ref{tab:cgshop-gaps} shows the gap between the best solution and the best bound that was found by CPLEX. In accordance with CPLEX, we use $gap(obj,b)=\frac{|b-obj|}{10^{-10}+|obj|}$ for calculating the gap, where $b$ is the best bound and $obj$ is the objective function value of the best integer solution. The table also includes the gap to the size of the convex hull for comparison. The edge-based approach was able to prove optimality for all instances of size $\leq25$. Most gaps between the convex hull and the best integer solution are below $0.10$ for larger instances. Despite the absolute differences, which are not accounted for by the specified bounds, the relative differences are quite small. The largest instance that was proven to be optimal was the \emph{euro-night-000045} instance, which contains $45$ points. To the best of our knowledge, this is the largest \textsc{Max-Area} instance that was solved to provable optimality. For instance sizes $>45$ we could not observe much improvement in the upper bounds in comparison to the trivial bound. One needs to add better cuts during the solving process to gather better bounds for these sizes.

  {
    \footnotesize
    \begin{longtable}{lcrrrrrr}
\toprule
           instance & solver & size & competition & IP &  best bound & gap & runtime (s) \\
\midrule
\textbf{euro-night-0000020} &      T &   20 &      6703352 &     6703352 &           6703352 &       0 &    5.0056 \\
    \textbf{london-0000020} &      T &   20 &     40978058 &    40978058 &          40978058 &       0 &    5.0056 \\
     \textbf{stars-0000020} &      T &   20 &  41041275182 & 41041275182 &       41041275182 &       0 &     13.29 \\
 \textbf{uniform-0000020-1} &      T &   20 &       188242 &      188242 &            188242 &       0 &   25.0389 \\
 \textbf{uniform-0000020-2} &      T &   20 &       130478 &      130478 &            130478 &       0 &       5.8 \\
  \textbf{us-night-0000020} &      T &   20 &      6615280 &     6615280 &           6615280 &       0 &      4.03 \\
\textbf{euro-night-0000025} &      T &   25 &      6235066 &     6235066 &           6235066 &       0 &   1954.56 \\
    \textbf{london-0000025} &      T &   25 &     31936238 &    31936238 &          31936238 &       0 &     37.03 \\
     \textbf{stars-0000025} &      T &   25 &  35893175226 & 35893175226 &       35893175226 &       0 &  582.5526 \\
          uniform-0000025-1 &      T &   25 &       319974 &      319974 &      307238.12120 & 0.03980 &         - \\
 \textbf{uniform-0000025-2} &      T &   25 &       351446 &      351446 &            351446 &       0 &  845.3083 \\
  \textbf{us-night-0000025} &      T &   25 &      9123288 &     9123288 &           9123288 &       0 &     73.08 \\
         euro-night-0000030 &      T &   30 &      7190308 &     7190308 &     3896089.33430 & 0.45810 &         - \\
             london-0000030 &      T &   30 &     15533240 &    15533240 &     9782667.47350 & 0.37020 &         - \\
              stars-0000030 &      T &   30 &  28318464852 & 28318464852 & 18121788612.62220 & 0.36010 &         - \\
          uniform-0000030-1 &      T &   30 &       373510 &      373510 &            353361 & 0.05390 &         - \\
          uniform-0000030-2 &      T &   30 &       427002 &      427002 &      283974.11740 & 0.33500 &         - \\
           us-night-0000030 &      T &   30 &      5750296 &     5750296 &     3965540.66670 & 0.31040 &         - \\
         euro-night-0000035 &      T &   35 &      5470084 &     5470084 &     3040936.87770 & 0.44410 &         - \\
    \textbf{london-0000035} &      E &   35 &     25363958 &    25363958 &          25363958 &       0 &    544.42 \\
              stars-0000035 &      T &   35 &  33183160522 & 33183160522 & 17014719428.44230 & 0.48720 &         - \\
          uniform-0000035-1 &      T &   35 &       499776 &      499776 &      238522.74400 & 0.52270 &         - \\
          uniform-0000035-2 &      T &   35 &       430856 &      430856 &      269129.42060 & 0.37540 &         - \\
           us-night-0000035 &      T &   35 &     10123092 &    10123092 &     4883885.67640 & 0.51760 &         - \\
\textbf{euro-night-0000040} &      E &   40 &      4423466 &     4423466 &           4423466 &       0 &  694.9955 \\
             london-0000040 &      E &   40 &     35905290 &    35905290 &                 1 &       1 &         - \\
              stars-0000040 &      E &   40 &  34302635012 & 34302635012 &                 1 &       1 &         - \\
          uniform-0000040-1 &      E &   40 &       777956 &      777956 &       18319.05880 & 0.97650 &         - \\
          uniform-0000040-2 &      E &   40 &       626084 &      626084 &                 1 &       1 &         - \\
           us-night-0000040 &      E &   40 &      5267414 &     5267414 &                 1 &       1 &         - \\
         euro-night-0000045 &      E &   45 &      5712456 &     5712456 &                 1 &       1 &         - \\
    \textbf{london-0000045} &      E &   45 &     33322976 &    33322976 &          33322976 &       0 &   39849.8 \\
              stars-0000045 &      E &   45 &  31524828442 & 31524828442 &                 1 &       1 &         - \\
          uniform-0000045-1 &      E &   45 &       813802 &      813802 &                 1 &       1 &         - \\
          uniform-0000045-2 &      E &   45 &       741648 &      741648 &                 1 &       1 &         - \\
           us-night-0000045 &      E &   45 &      3595152 &     3595152 &             95814 & 0.97330 &         - \\
         euro-night-0000050 &      E &   50 &      7204726 &     7204726 &                 1 &       1 &         - \\
             london-0000050 &      E &   50 &     31064970 &    31064970 &                 1 &       1 &         - \\
     \textbf{stars-0000050} &      E &   50 &  26625487604 & 26625487604 &       26625487604 &       0 &   7796.55 \\
          uniform-0000050-1 &      E &   50 &       625044 &      625044 &                 1 &       1 &         - \\
 \textbf{uniform-0000050-2} &      E &   50 &      1094266 &     1094266 &           1094266 &       0 &   7903.57 \\
           us-night-0000050 &      E &   50 &      5054470 &     5054470 &                 1 &       1 &         - \\
\bottomrule
\caption{CG:SHOP results for \textsc{Min-Area} (both approaches) for instances of size $20-50$}
\label{tab:cgshop-gaps-min}
\end{longtable}

  }

For the edge-based approach the minimization variant is significantly harder than the \textsc{Max-Area} problem for most instances. Unfortunately, the triangulation-based approach consumes a lot of space on larger point sets. Table~\ref{tab:cgshop-gaps-min} summarizes our results on \textsc{Min-Area}. For instances of size $20-35$ (except \emph{london-0000035}), the best results could be obtained using \textsc{TriangulationV1}. For the other instances the space requirements were too high and thus the edge-based approach was used. Apart from the instance \emph{uniform-0000025-1} all instances of size $20-25$ could be solved to optimality.
For most of the larger instances of size $>35$, the edge-based approach was unable to improve the trivial bound of 1 (all solutions in the competition must have integral area). However, the competition results for the instances \emph{london-0000045}, \emph{uniform-0000050-2} and \emph{stars-0000050} were proven to be optimal. To the best of our knowledge \emph{uniform-0000050-2} and \emph{stars-0000050} are the largest \textsc{Min-Area} instances that could be solved optimally. For instance size $>50$ we did not observe any improvement in the bounds. Apart from the optimal solutions, the edge-based approach is not well-suited for the minimization variant. Other approaches or improvements to the triangulation-based IP will most likely help to find better bounds.

	\section{Conclusions}
\label{sec:conclusions}

While our work shows that with some amount of algorithm engineering,
it is possible to extend the range of instances that can be solved to
provable optimality, it also illustrates the practical difficulty of the problem.
This reflects the limitations of such IP-based methods: The edge-based approach
makes use of an asymmetric variant of the TSP, which is known to be harder than the symmetric TSP,
while the triangle-based approach suffers from its inherently large number of variable and constraints.
Furthermore, the non-local nature of \textsc{Min-Area} and \textsc{Max-Area} polygons
(which may contain edges that connect far-away points) makes it difficult to reduce
the set of candidate edges. 

As a result, \textsc{Min-Area} and \textsc{Max-Area}
turn out to be prototypes of geometric optimization problems that are difficult
both in theory and practice. This differs fundamentally from a problem such as 
\textsc{Minimum Weight Triangulation}, for which provably optimal solutions 
to huge point sets can be found~\cite{mwt1}, and practically difficult instances
seem elusive~\cite{mwt2}

\section*{Acknowledgments}
This work was supported by DFG grant FE407/21-1, ``Computational Geometry: Solving Hard Optimization Problems (CG:SHOP)''.
We thank an anonymous reviewer for various constructive comments that helped to improve the overall presentation.

\bibliographystyle{ACM-Reference-Format}
\bibliography{bibliography}


\begin{thebibliography}{14}


\ifx \showCODEN    \undefined \def \showCODEN     #1{\unskip}     \fi
\ifx \showDOI      \undefined \def \showDOI       #1{#1}\fi
\ifx \showISBNx    \undefined \def \showISBNx     #1{\unskip}     \fi
\ifx \showISBNxiii \undefined \def \showISBNxiii  #1{\unskip}     \fi
\ifx \showISSN     \undefined \def \showISSN      #1{\unskip}     \fi
\ifx \showLCCN     \undefined \def \showLCCN      #1{\unskip}     \fi
\ifx \shownote     \undefined \def \shownote      #1{#1}          \fi
\ifx \showarticletitle \undefined \def \showarticletitle #1{#1}   \fi
\ifx \showURL      \undefined \def \showURL       {\relax}        \fi
\providecommand\bibfield[2]{#2}
\providecommand\bibinfo[2]{#2}
\providecommand\natexlab[1]{#1}
\providecommand\showeprint[2][]{arXiv:#2}

\bibitem[\protect\citeauthoryear{Demaine, Fekete, Keldenich, Krupke, and
  Mitchell}{Demaine et~al\mbox{.}}{2022}]%
        {challenge19}
\bibfield{author}{\bibinfo{person}{Erik~D. Demaine},
  \bibinfo{person}{S\'andor~P. Fekete}, \bibinfo{person}{Phillip Keldenich},
  \bibinfo{person}{Dominik Krupke}, {and} \bibinfo{person}{Joseph~S.B.
  Mitchell}.} \bibinfo{year}{2022}\natexlab{}.
\newblock \showarticletitle{Area-Optimal Polygonalizations: The 2019 {CG
  Challenge}}.
\newblock \bibinfo{journal}{\emph{Journal of Experimental Algorithms}}
  (\bibinfo{year}{2022}).
\newblock
\newblock
\shownote{To appear.}


\bibitem[\protect\citeauthoryear{Fekete}{Fekete}{1992}]%
        {f-gtsp-92}
\bibfield{author}{\bibinfo{person}{S{\'a}ndor~P. Fekete}.}
  \bibinfo{year}{1992}\natexlab{}.
\newblock \emph{\bibinfo{title}{Geometry and the Travelling Salesman Problem}}.
\newblock Ph.{D}. Thesis. \bibinfo{school}{Department of Combinatorics and
  Optimization, University of Waterloo}, \bibinfo{address}{Waterloo, ON}.
\newblock


\bibitem[\protect\citeauthoryear{Fekete}{Fekete}{2000}]%
        {fekete2000simple}
\bibfield{author}{\bibinfo{person}{S{\'a}ndor~P. Fekete}.}
  \bibinfo{year}{2000}\natexlab{}.
\newblock \showarticletitle{On simple polygonizations with optimal area}.
\newblock \bibinfo{journal}{\emph{Discrete \& Computational Geometry}}
  \bibinfo{volume}{23}, \bibinfo{number}{1} (\bibinfo{year}{2000}),
  \bibinfo{pages}{73--110}.
\newblock


\bibitem[\protect\citeauthoryear{Fekete, Friedrichs, Hemmer, Papenberg,
  Schmidt, and Troegel}{Fekete et~al\mbox{.}}{2015}]%
        {fekete2015area}
\bibfield{author}{\bibinfo{person}{S{\'a}ndor~P. Fekete},
  \bibinfo{person}{Stephan Friedrichs}, \bibinfo{person}{Michael Hemmer},
  \bibinfo{person}{Melanie Papenberg}, \bibinfo{person}{Arne Schmidt}, {and}
  \bibinfo{person}{Julian Troegel}.} \bibinfo{year}{2015}\natexlab{}.
\newblock \showarticletitle{Area-and boundary-optimal polygonalization of
  planar point sets}. In \bibinfo{booktitle}{\emph{European Workshop on
  Computational Geometry (EuroCG)}}. \bibinfo{pages}{133--136}.
\newblock


\bibitem[\protect\citeauthoryear{Fekete, Haas, Krupke, Lieder, Niehs, Perk,
  Sack, and Scheffer}{Fekete et~al\mbox{.}}{2020}]%
        {mwt2}
\bibfield{author}{\bibinfo{person}{S\'andor~P. Fekete},
  \bibinfo{person}{Andreas Haas}, \bibinfo{person}{Dominik Krupke},
  \bibinfo{person}{Yannic Lieder}, \bibinfo{person}{Eike Niehs},
  \bibinfo{person}{Michael Perk}, \bibinfo{person}{Victoria Sack}, {and}
  \bibinfo{person}{Christian Scheffer}.} \bibinfo{year}{2020}\natexlab{}.
\newblock \showarticletitle{Hard Instances of the Minimum-Weight Triangulation
  Problem}. In \bibinfo{booktitle}{\emph{European Workshop on Computational
  Geometry (EuroCG)}}. \bibinfo{pages}{29:1--29:9}.
\newblock


\bibitem[\protect\citeauthoryear{Fekete and Pulleyblank}{Fekete and
  Pulleyblank}{1993}]%
        {fekete1993area}
\bibfield{author}{\bibinfo{person}{S{\'a}ndor~P. Fekete} {and}
  \bibinfo{person}{William~R. Pulleyblank}.} \bibinfo{year}{1993}\natexlab{}.
\newblock \showarticletitle{Area optimization of simple polygons}. In
  \bibinfo{booktitle}{\emph{Proc. 9th Symposium on Computational Geometry
  (SoCG)}}. \bibinfo{pages}{173--182}.
\newblock


\bibitem[\protect\citeauthoryear{Ford and Fulkerson}{Ford and
  Fulkerson}{1956}]%
        {ford1956maximal}
\bibfield{author}{\bibinfo{person}{Lester~R Ford} {and}
  \bibinfo{person}{Delbert~R Fulkerson}.} \bibinfo{year}{1956}\natexlab{}.
\newblock \showarticletitle{Maximal flow through a network}.
\newblock \bibinfo{journal}{\emph{Canadian journal of Mathematics}}
  \bibinfo{volume}{8}, \bibinfo{number}{3} (\bibinfo{year}{1956}),
  \bibinfo{pages}{399--404}.
\newblock


\bibitem[\protect\citeauthoryear{Gomory and Hu}{Gomory and Hu}{1961}]%
        {gomory1961multi}
\bibfield{author}{\bibinfo{person}{Ralph~E Gomory} {and}
  \bibinfo{person}{Tien~Chung Hu}.} \bibinfo{year}{1961}\natexlab{}.
\newblock \showarticletitle{Multi-terminal network flows}.
\newblock \bibinfo{journal}{\emph{J. Soc. Indust. Appl. Math.}}
  \bibinfo{volume}{9}, \bibinfo{number}{4} (\bibinfo{year}{1961}),
  \bibinfo{pages}{551--570}.
\newblock


\bibitem[\protect\citeauthoryear{Haas}{Haas}{2018}]%
        {mwt1}
\bibfield{author}{\bibinfo{person}{Andreas Haas}.}
  \bibinfo{year}{2018}\natexlab{}.
\newblock \showarticletitle{Solving Large-Scale Minimum-Weight Triangulation
  Instances to Provable Optimality}. In \bibinfo{booktitle}{\emph{Symposium on
  Computational Geometry (SoCG)}}. \bibinfo{pages}{44:1--44:14}.
\newblock


\bibitem[\protect\citeauthoryear{Papenberg}{Papenberg}{2014}]%
        {Papenberg2014}
\bibfield{author}{\bibinfo{person}{Melanie Papenberg}.}
  \bibinfo{year}{2014}\natexlab{}.
\newblock \emph{\bibinfo{title}{{Exact methods for area-optimal polygons}}}.
\newblock \bibinfo{thesistype}{Master's\ thesis}. \bibinfo{school}{Braunschweig
  University of Technology}.
\newblock


\bibitem[\protect\citeauthoryear{Peethambaran, Parakkat, and
  Muthuganapathy}{Peethambaran et~al\mbox{.}}{2015}]%
        {peethambaran2015randomized}
\bibfield{author}{\bibinfo{person}{Jiju Peethambaran},
  \bibinfo{person}{Amal~Dev Parakkat}, {and} \bibinfo{person}{Ramanathan
  Muthuganapathy}.} \bibinfo{year}{2015}\natexlab{}.
\newblock \showarticletitle{A randomized approach to volume constrained
  polyhedronization problem}.
\newblock \bibinfo{journal}{\emph{Journal of Computing and Information Science
  in Engineering}} \bibinfo{volume}{15}, \bibinfo{number}{1}
  (\bibinfo{year}{2015}), \bibinfo{pages}{011009}.
\newblock


\bibitem[\protect\citeauthoryear{Peethambaran, Parakkat, and
  Muthuganapathy}{Peethambaran et~al\mbox{.}}{2016}]%
        {peethambaran2016empirical}
\bibfield{author}{\bibinfo{person}{Jiju Peethambaran},
  \bibinfo{person}{Amal~Dev Parakkat}, {and} \bibinfo{person}{Ramanathan
  Muthuganapathy}.} \bibinfo{year}{2016}\natexlab{}.
\newblock \showarticletitle{An Empirical Study on Randomized Optimal Area
  Polygonization of Planar Point Sets}.
\newblock \bibinfo{journal}{\emph{Journal of Experimental Algorithmics}}
  \bibinfo{volume}{21} (\bibinfo{year}{2016}), \bibinfo{pages}{1--10}.
\newblock


\bibitem[\protect\citeauthoryear{Stoer and Wagner}{Stoer and Wagner}{1997}]%
        {stoer1997simple}
\bibfield{author}{\bibinfo{person}{Mechthild Stoer} {and}
  \bibinfo{person}{Frank Wagner}.} \bibinfo{year}{1997}\natexlab{}.
\newblock \showarticletitle{A simple min-cut algorithm}.
\newblock \bibinfo{journal}{\emph{Journal of the ACM (JACM)}}
  \bibinfo{volume}{44}, \bibinfo{number}{4} (\bibinfo{year}{1997}),
  \bibinfo{pages}{585--591}.
\newblock


\bibitem[\protect\citeauthoryear{Taranilla, Gagliardi, and
  Hern{\'a}ndez~Pe{\~n}alver}{Taranilla et~al\mbox{.}}{2011}]%
        {taranilla2011approaching}
\bibfield{author}{\bibinfo{person}{Maria~Teresa Taranilla},
  \bibinfo{person}{Edilma~Olinda Gagliardi}, {and} \bibinfo{person}{Gregorio
  Hern{\'a}ndez~Pe{\~n}alver}.} \bibinfo{year}{2011}\natexlab{}.
\newblock \showarticletitle{Approaching minimum area polygonization}. In
  \bibinfo{booktitle}{\emph{Congreso Argentino de Ciencias de la Computaci\'on
  (CACIC)}}. \bibinfo{pages}{161--170}.
\newblock


\end{thebibliography}

\end{document}